\documentclass[prd,reprint,showpacs,superscriptaddress,bibnotes,floatfix]{revtex4-1}
\usepackage{natbib}
\usepackage{graphics}
\usepackage{multirow}
\usepackage{amsbsy}
\usepackage{mathrsfs}
\usepackage{footmisc}
\usepackage{hyperref}
\hypersetup{
  colorlinks = true,
  urlcolor = black,
  linkcolor=black,
  citecolor=black
}
\def\act{ACT} 

\def\bicepone{BICEP1}
\def\biceptwo{BICEP2}
\def\bicepthree{BICEP3}

\def\keck{{\it Keck}}
\def\keckarray{{\it Keck Array}}
\def\planck{{\it Planck}} 

\def\wmap{WMAP}
\def\spt{SPT}

\def\polarbear{POLARBEAR}

\def\healpix{{\sc healp}ix}

\def\muK{~\mu{\rm K}}

\def\deg{^\circ}
\def\emode{$E$-mode}
\def\bmode{$B$-mode}

\def\lcdm{$\Lambda$CDM}

\def\bI{BK-I}
\def\bV{BK-V}
\def\piXXX{PIP-XXX}

\def\sovast{Sov.\ Astron.}
\def\aap{Astron.\ Astrophys.}

\def\apj{Astrophys.\ J.}

\def\apjs{Astrophys.\ J.\ Suppl.\ Ser.}

\def\jcap{J.\ Cosmol.\ Astropart.\ Phys.}

\def\mnras{Mon.\ Not.\ R.\ Astron.\ Soc.}

\def\prd{Phys.\ Rev.\ D}

\def\prl{Phys.\ Rev.\ Lett.}

\begin{document}

\title{A Joint Analysis of \biceptwo/\keckarray\ and \planck\ Data}

\author{BICEP2/Keck and Planck Collaborations: P.~A.~R.~Ade}
\affiliation{School of Physics and Astronomy, Cardiff University, Queens Buildings, The Parade, Cardiff, CF24 3AA, U.K.}
\author{N.~Aghanim}
\affiliation{Institut d'Astrophysique Spatiale, CNRS (UMR8617) Universit\'{e} Paris-Sud 11, B\^{a}timent 121, Orsay, France}
\author{Z.~Ahmed}
\affiliation{Department of Physics, Stanford University, Stanford, California 94305, U.S.A.}
\author{R.~W.~Aikin}
\affiliation{California Institute of Technology, Pasadena, California, U.S.A.}
\author{K.~D.~Alexander}
\affiliation{Harvard-Smithsonian Center for Astrophysics, 60 Garden Street MS 42, Cambridge, Massachusetts 02138, U.S.A.}
\author{M.~Arnaud}
\affiliation{Laboratoire AIM, IRFU/Service d'Astrophysique - CEA/DSM - CNRS - Universit\'{e} Paris Diderot, B\^{a}t. 709, CEA-Saclay, F-91191 Gif-sur-Yvette Cedex, France}
\author{J.~Aumont}
\affiliation{Institut d'Astrophysique Spatiale, CNRS (UMR8617) Universit\'{e} Paris-Sud 11, B\^{a}timent 121, Orsay, France}
\author{C.~Baccigalupi}
\affiliation{SISSA, Astrophysics Sector, via Bonomea 265, 34136, Trieste, Italy}
\author{A.~J.~Banday}
\affiliation{Universit\'{e} de Toulouse, UPS-OMP, IRAP, F-31028 Toulouse cedex 4, France}
\affiliation{CNRS, IRAP, 9 Av. colonel Roche, BP 44346, F-31028 Toulouse cedex 4, France}
\author{D.~Barkats}
\affiliation{Joint ALMA Observatory, Vitacura, Santiago, Chile}
% ...

\date[Draft~]{\today}

\begin{abstract}
We report the results of a joint analysis of data from \biceptwo/\keckarray\
and \planck.
\biceptwo\ and \keckarray\ have observed the same approximately $400$ deg$^2$
patch of sky centered on RA 0h, Dec.\ $-57.5\deg$.
The combined maps reach a depth of 57\,nK\,deg in Stokes $Q$ and $U$
in a band centered at 150\,GHz.
\planck\ has observed the full sky in polarization
at seven frequencies from 30 to 353\,GHz, but much less deeply
in any given region (1.2\,$\mu$K\,deg in $Q$ and $U$ at 143\,GHz).
We detect 150$\times$353 cross-correlation in {\bmode}s at high
significance.
We fit the single- and cross-frequency power spectra at frequencies 
$\geq 150$\,GHz to a lensed-\lcdm\ model that includes dust and a possible
contribution from inflationary gravitational waves 
(as parameterized by the tensor-to-scalar ratio $r$),
using a prior on the frequency spectral behavior
of polarized dust emission from previous \planck\ analysis of
other regions of the sky.
We find strong evidence for dust and no statistically
significant evidence for tensor modes.
We probe various model variations and extensions, including
adding a synchrotron component in combination with lower
frequency data, and find that these make little
difference to the $r$ constraint.
Finally we present an alternative analysis which is similar
to a map-based cleaning of the dust contribution, and show
that this gives similar constraints.
The final result is expressed as a likelihood 
curve for $r$, and yields an upper limit $r_{0.05}<0.12$ at 95\% confidence.
Marginalizing over dust and $r$, lensing
{\bmode}s are detected at $7.0\,\sigma$ significance.
\end{abstract}

\doi{10.1103/PhysRevLett.114.101301}

\maketitle

\section{Introduction}

\setcounter{footnote}{1}

The cosmic microwave background (CMB)~\cite{penzias65}, is an essential source
of information about all epochs of the Universe.  
In the past several decades, characterization of the temperature and polarization
anisotropies of the CMB has helped to establish the standard
cosmological model (\lcdm) and to measure its parameters to high
precision (see for example Refs.~\cite{bennett13,planck2013XVI}).

An extension to the standard big bang model, inflation, postulates a short
period of exponential expansion in the very early Universe, naturally
setting the initial conditions required by 
\lcdm, as well as solving a number of additional problems in standard
cosmology.
Inflation's basic predictions regarding the Universe’s large-scale
geometry and structure have been borne out by cosmological
measurements to date (see Ref.~\cite{planck2013XXII} for a review).  
Inflation makes an additional prediction, the existence
of a background of gravitational waves, or tensor mode
perturbations~\cite{starobinsky79,rubakov82,fabbri83,abbott84}.
At the recombination epoch, the inflationary gravitational waves
(IGW) contribute to the anisotropy of the CMB in both total intensity and
linear polarization. 
The amplitude of tensors is conventionally parameterized by $r$, the
tensor-to-scalar ratio at a fiducial scale.
Theoretical predictions of the value of $r$ cover a very wide range. 
Conversely, a measurement of $r$ can discriminate between models of inflation. 

Tensor modes produce a small increment in the temperature anisotropy
power spectrum over the standard \lcdm\ scalar perturbations at
multipoles $\ell \lesssim 60$; measuring this increment requires the large sky
coverage traditionally achieved by space-based experiments,
and an understanding of the other cosmological parameters.
The effects of tensor perturbations on \bmode\ polarization is less ambiguous
than on temperature or \emode\ polarization over the range $\ell
\lesssim 150$. 
The \bmode\ polarization signal produced by scalar
perturbations is very small and is dominated by the weak lensing of \emode\
polarization on small angular scales, making the detection
of an IGW contribution possible~\cite{polnarev85,seljak97a,seljak97b,kamionkowski97}.

\planck~\footnote{\planck\ (\url{http://www.esa.int/planck}) is a
  project of the European Space Agency (ESA) with instruments provided
  by two scientific consortia funded by ESA member states (in
  particular the lead countries, France and Italy) with contributions
  from NASA (USA), and telescope reflectors provided in a
  collaboration between ESA and a scientific consortium led and funded
  by Denmark.} was the third generation CMB space mission, which
mapped the full sky in polarization in seven bands centered at frequencies
from 30\,GHz to 353\,GHz to a resolution of 33 to
5\,arcminutes~\cite{planck2011I,planck2013I}.   
The \planck\ collaboration has published the best limit
to date on tensor modes using CMB data alone~\cite{planck2013XVI}:
$r_{0.002}<0.11$ (at 95\% confidence) using a combination
of \planck, \spt\ and \act\ temperature data, plus \wmap\
polarization, although the \planck\ $r$ limit is model-dependent, with
running of the scalar spectral index or additional relativistic
degrees of freedom being well-known degeneracies which allow larger
values of $r$.

Interstellar dust grains produce thermal emission,
the brightness of which increases rapidly from the
100--150~GHz frequencies favored for CMB observations, becoming
dominant at $\geq 350$~GHz even at high galactic latitude.
The dust grains align with the Galactic magnetic field to
produce emission with a degree of linear polarization~\cite{hildebrand99}.
The observed degree of polarization depends on the structure of the Galactic
magnetic field along the line of sight, as well as the properties of
the dust grains (see for example Refs.~\cite{draine04,martin07}).
This polarized dust emission results in both \emode\ and \bmode,
and acts as a potential contaminant to a measurement of $r$.
Galactic dust polarization was detected by
Archeops~\cite{archeopsdust} at 353\,GHz and by \wmap~\cite{page07,bennett13}
at 90\,GHz.

\biceptwo\ was a specialized, low angular resolution experiment,
which operated from the South Pole from 2010 to 2012, concentrating
150\,GHz sensitivity comparable to \planck\ on a roughly 1\,\% patch of
sky at high Galactic latitude~\cite{b2instpap14}.
The \biceptwo\ Collaboration published a highly
significant detection of \bmode\ polarization in excess of the $r$=0 
lensed-\lcdm\ expectation over the range $30<\ell<150$
in Ref.~\cite[hereafter \bI]{biceptwoI}.
Modest evidence against a thermal Galactic dust component
dominating the observed signal was presented based on the
cross-spectrum against 100\,GHz maps from the previous \bicepone\ 
experiment.
The detected \bmode\ level was higher than that projected
by several existing dust models~\cite{dunkley08,delabrouille13}
although these did not claim any high degree of reliability.

The \planck\ survey released information on the structure of the
dust polarization sky at intermediate latitudes~\cite{planckiXIX}, and the
frequency dependence of the polarized dust emission at frequencies
relevant to CMB studies~\cite{planckiXXII}.
Other papers argued that the \biceptwo\ region is
significantly contaminated by dust~\cite{flauger14,mortonson14}.
Finally \planck\ released information on
dust polarization at high latitude~\cite[hereafter \piXXX]{planckiXXX},
and in particular examined a field
centered on the \biceptwo\ region (but somewhat larger than it)
finding a level of polarized dust emission at 353\,GHz sufficient
to explain the 150\,GHz excess observed by \biceptwo, although
with relatively low signal-to-noise.

\keckarray\ is a system of \biceptwo-like receivers
also located at the South Pole.
During the 2012 and 2013 seasons \keckarray\ observed
the same field as \biceptwo\ in the same 150\,GHz
frequency band~\cite[hereafter \bV]{biceptwoV}.
Combining the \biceptwo\ and \keckarray\ maps
yields $Q$ and $U$ maps with rms noise of 57\,nK in
nominal $1\,\mathrm{deg}^2$  pixels---by far the
deepest made to date.

In this paper, we take cross-spectra between the joint
\biceptwo/\keck\ maps and all the polarized bands of \planck.
The structure is as follows.
In~Sec.~\ref{sec:data} we describe the preparation of
the input maps, the expectations for dust, and the
power spectrum results.
In Sec.~\ref{sec:likeanal} the main multi-frequency
cross-spectrum likelihood method is introduced
and applied to the data, and a number of variations
from the selected fiducial analysis are explored.
Sec.~\ref{sec:valid} describes validation tests
using simulations as well as an alternate likelihood.
In Sec.~\ref{sec:decorr} we investigate whether there
could be decorrelation between the \planck\
and \biceptwo/\keck\ maps due to the astrophysics
of dust and/or instrumental effects.
Finally we conclude in Sec.~\ref{sec:conc}.

\section{Maps to power spectra}
\label{sec:data}

\subsection{Maps and preparation}

We primarily use the \biceptwo/\keck\ combined maps, as described
in \bV.
We also use the \biceptwo-only and \keck-only maps as a cross check.
The \planck\ maps used for cross-correlation with
\biceptwo/\keck\ are the full-mission polarized maps from the 
PR2 \planck\ science release~\footnote{\url{http://archives.esac.esa.int/pla2}}~\cite{planck2014-a01},
a subset of which was presented in \piXXX.
We compute \planck\ single-frequency spectra as the cross-power spectra
of two data-split maps, in which the data are split into two
subsets with independent noise.
We consider three data split maps: (i) detector-set maps, where the detectors
at a given frequency are divided into two groups, (ii) yearly maps, where the data
from the first and second years of observations are used for the two
maps, and (iii) half-ring maps, where the data from each pointing period
is divided in halves.
To evaluate uncertainties due to \planck\ instrumental noise, we use
500 noise simulations of each map; these are the standard set
of time-ordered data noise simulations projected into 
sky maps (the FFP8 simulations defined in Ref.~\cite{planck2014-a14}).

While the \planck\ maps are filtered only by the
instrument beam (the effective beam defined in Refs.~\cite{planck2013IV} and
\cite{planck2013VII}), the \biceptwo/\keck\ maps 
are in addition filtered due to the observation
strategy and analysis process.
In particular, large angular scales are suppressed anisotropically in
the \biceptwo/\keck\ mapmaking process to avoid atmospheric and ground-fixed
contamination; this suppression is corrected in the power spectrum
estimate. 
In order to facilitate comparison, we therefore prepare
``\planck\ as seen by \biceptwo/\keck'' maps.
In the first step we use the {\sc anafast}, {\sc alteralm} and {\sc synfast} routines
from the \healpix~\footnote{
\url{http://healpix.sourceforge.net/}}
package~\cite{healpix} to resmooth the \planck\ maps with the 
\biceptwo/\keck\ beam profile, assuming azimuthal symmetry of the beam.
The coordinate rotation from Galactic to celestial coordinates of the
$T$, $Q$, and $U$ maps is performed using the {\sc alteralm} routine in the
\healpix\ package. 
The sign of the Stokes $U$ map is flipped to convert from the \healpix\
to the IAU polarization convention.
Next we pass these through the ``observing'' matrix $R$, described
in~Section~VI.B of \bI, to produce maps
that include the filtering of modes occuring in the
data processing
pipeline (including polynominal filtering and scan-synchronous
template removal, plus deprojection of beam systematics).

\begin{figure*}
\resizebox{\textwidth}{!}{\includegraphics{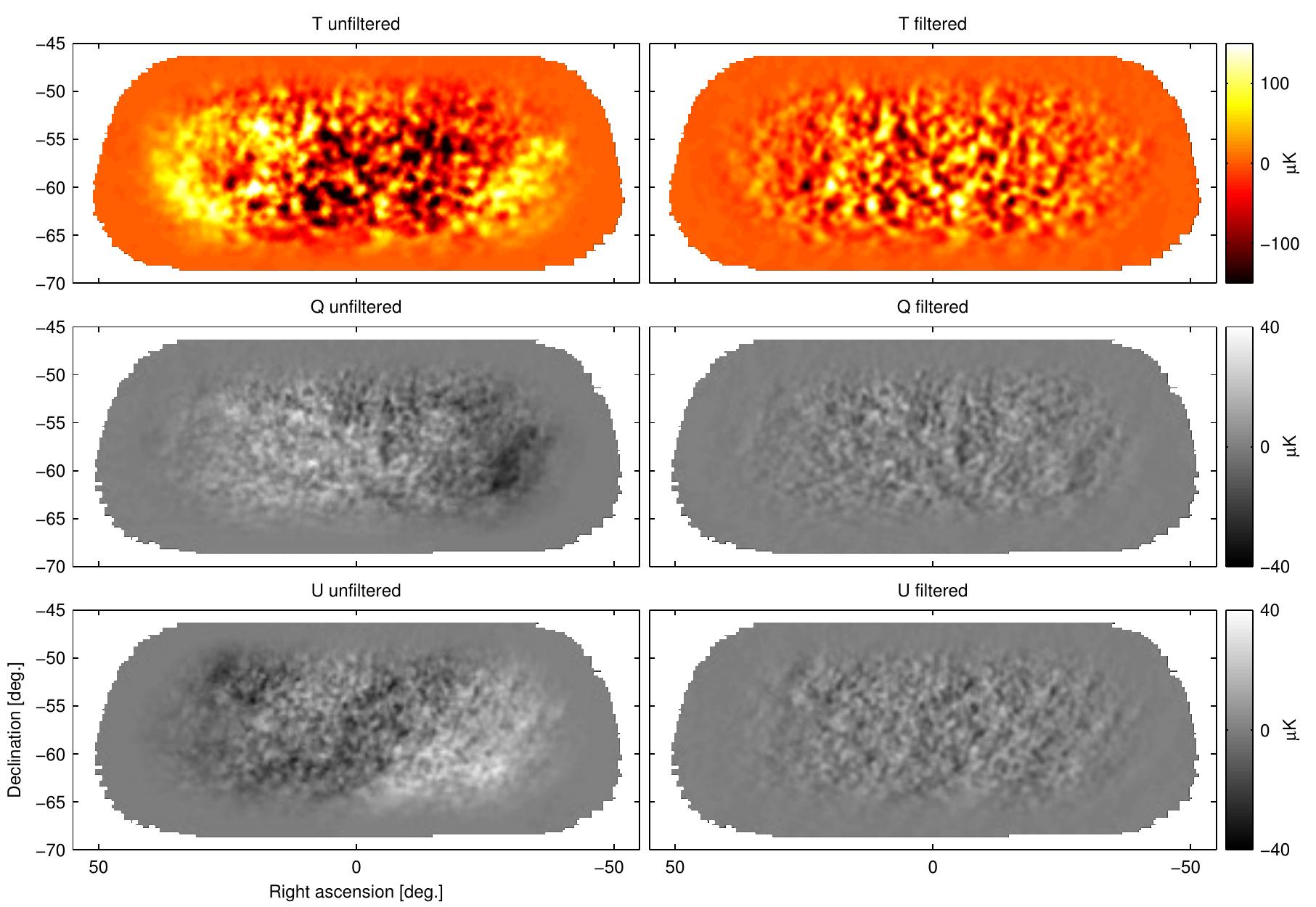}}
\caption{\planck\ 353\,GHz $T$, $Q$, and $U$ maps before
(left) and after (right) the application of \biceptwo/\keck\ filtering.
In both cases the maps have been multiplied by the \biceptwo/\keck\
apodization mask.
The \planck\ maps are presmoothed to the \biceptwo/\keck\
beam profile and have the mean value subtracted.
The filtering, in particular the third order polynominal subtraction
to suppress atmospheric pickup, removes large-angular scale signal along the 
\biceptwo/\keck\ scanning direction (parallel to the right ascension
direction in the maps here).}
\label{fig:planck353maps}
\end{figure*}

Figure~\ref{fig:planck353maps} shows the resmoothed \planck\
353\,GHz $T$, $Q$, and $U$ maps before and after filtering.
In both cases the \biceptwo/\keck\ inverse variance
apodization mask has been applied.
This figure emphasizes the need to account for the filtering
before any comparison of maps is attempted,
either qualitative or quantitative.

\subsection{Expected spatial and frequency spectra of dust} 
\label{sec:dustexpect}

Before examining the power spectra it is useful to review
expectations for the spatial and frequency spectra
of dust.
Figure~2 of \piXXX\ shows that the dust $BB$ (and $EE$) angular
power spectra are well fit by a simple power law
$\mathcal{D}_\ell \propto \ell^{-0.42}$, where $\mathcal{D}_\ell =
C_\ell \ell \left(\ell + 1\right) / 2\pi$,
when averaging over large regions of sky outside of the Galactic
plane.
Section 5.2 of the same paper states that there is no
evidence for departure from this behavior for 1\% sky
patches, although the signal-to-noise ratio is low for some regions.
Presumably we expect greater 
fluctuation from the mean behavior than would be
expected for a Gaussian random field.

The spectral energy distribution (SED) of dust polarization 
was measured in Ref.~\cite{planckiXXII} for 400 patches with
$10\deg$ radius at intermediate Galactic latitudes.
Figure~10 of this reference shows emprically that the
mean polarized dust SED is described by a simple modified blackbody spectrum with 
$T_\mathrm{d}=19.6$\,K and $\beta_\mathrm{d}=1.59 \pm 0.17$
to within an accuracy of a few percent over the frequency
range 100--353\,GHz.
Within this frequency range variations in the two
parameters are highly degenerate and the
choice is made to hold $T_\mathrm{d}$ fixed at the value obtained
from a fit to the SED of total intensity, and describe any variation
with the $\beta_\mathrm{d}$ parameter.
The uncertainty on $\beta_\mathrm{d}$ quoted above is the $1\sigma$
dispersion of the individual patch measurements, and hence is 
an upper limit, since some of the fluctuation 
is due to noise rather than real variation on the sky.
This SED is confirmed to be a good match to data when 
averaging over 24\% of the cleanest high latitude sky in
Fig.~6 of \piXXX.

\subsection{Power spectrum estimation and results}

\begin{figure*}
\resizebox{\textwidth}{!}{\includegraphics{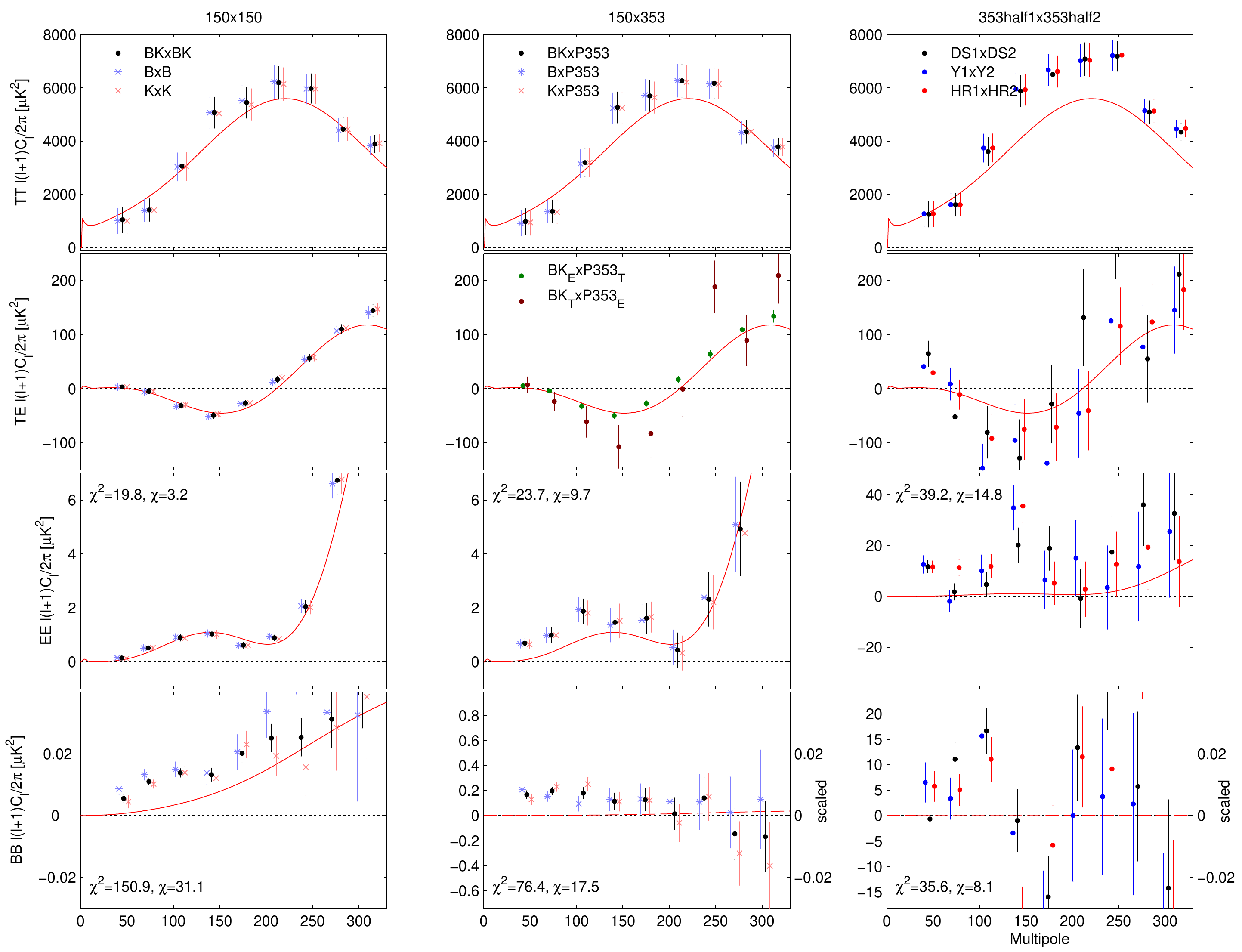}}
\caption{Single- and cross-frequency spectra between \biceptwo/\keck\
maps at 150\,GHz and \planck\ maps at 353\,GHz.
The red curves show the lensed-\lcdm\ expectations.
The left column shows single-frequency spectra of the
\biceptwo, \keckarray\ and combined \biceptwo/\keck\ maps.
The \biceptwo\ spectra are identical to those in \bI,
while the \keckarray\ and combined are as given in
\bV.
The center column shows cross-frequency spectra between \biceptwo/\keck\
maps and \planck\ 353\,GHz maps.
The right column shows \planck\ 353\,GHz data-split cross-spectra.
In all cases the error bars are the standard deviations of
lensed-\lcdm+noise simulations and hence contain no sample
variance on any other component.
For $EE$ and $BB$ the $\chi^2$ and $\chi$ (sum of deviations)
versus lensed-\lcdm\ for the nine bandpowers shown is marked at upper/lower left
(for the combined \biceptwo/\keck\ points and DS1$\times$DS2).
In the bottom row (for $BB$) the center and right panels have  a scaling applied such
that signal from dust with the fiducial frequency spectrum
would produce signal with the same apparent amplitude as in the 150\,GHz
panel on the left (as indicated by the right-side $y$-axes).
We see from the significant excess apparent in the bottom center panel 
that a substantial amount of the signal detected 
at 150\,GHz by \biceptwo\ and \keckarray\ indeed appears to be
due to dust.}

\label{fig:spectra}
\end{figure*}

The power spectrum estimation proceeds exactly as in
\bI, including the matrix based purification
operation to prevent $E$ to $B$ mixing.
Figure~\ref{fig:spectra} shows the results
for \biceptwo/\keck\ and \planck\ 353\,GHz for
$TT$, $TE$, $EE$, and $BB$.
In all cases the error bars are the standard deviations of
lensed-\lcdm+noise simulations~\footnote{
With parameters taken from \planck~\cite{planck2013XVI}.}
and hence contain no sample
variance on any other component.
The results in the left column are auto-spectra, identical
to those given in \bI\ and \bV---these spectra are consistent with
lensed-\lcdm+noise except for the excess in $BB$
for $\ell<200$.

The right column of Fig.~2 shows cross-spectra between two
halves of the \planck\ 353\,GHz data set, with three
different splits shown.
The \planck\ collaboration prefers the use of cross-spectra even at a
single frequency to gain additional immunity to
systematics and to avoid the need to noise
debias auto-spectra.
The $TT$ spectrum is higher than \lcdm\
around $\ell=200$---presumably due to a dust
contribution.
The $EE$ and $BB$ spectra are noisy, but both appear to show
an excess over \lcdm\ for $\ell<150$---again presumably due to dust.
We note that these spectra do not appear to follow the
power-law expectation mentioned in Sec.~\ref{sec:dustexpect},
but we emphasize that the error bars
contain no sample variance on any dust component
(Gaussian or otherwise).

The center column of Fig.~2 shows cross-spectra between \biceptwo/\keck\
and \planck\ maps.
For $TE$ one can use the $T$-modes from \biceptwo\ and the
$E$-modes from \planck\ or vice versa and both options are shown.
Since the $T$-modes are very similar between the two
experiments, these $TE$ spectra look similar to the
single-experiment $TE$ spectrum which shares the $E$-modes.
The $EE$ and $BB$ cross-spectra are the most interesting---there
appears to be a highly significant detection of correlated
\bmode\ power between 150 and 353\,GHz, with the pattern
being much brighter at 353, consistent with the expectation
from dust.
We also see hints of detection in the $EE$ spectrum---while
dust $E$-modes are subdominant to the cosmological signal
at 150\,GHz, the weak dust contribution enhances the BK150$\times$P353
cross-spectrum at $\ell\approx100$.

The polarized dust SED model mentioned in
Sec.~\ref{sec:dustexpect} implies
that dust emission is approximately 25 times brighter
in the \planck\ 353\,GHz band than it is in the
\biceptwo/\keck\ 150\,GHz band
(integrating appropriately over the instrumental bandpasses).
The expectation for a dust-dominated spectrum is thus that the
BK150$\times$P353 cross-spectrum should have an amplitude 25 times
that of BK150$\times$BK150, and P353$\times$P353 should be 25 times
higher again. The $y$-axis scaling in the bottom row
of Fig.~\ref{fig:spectra} has been adjusted so that
a dust signal obeying this rule will have equal apparent
amplitude in each panel.
We see that a substantial amount of the BK150$\times$BK150
signal indeed appears to be due to dust.

To make a rough estimate of the significance of deviation
from lensed-\lcdm, we calculate $\chi^2$ and $\chi$
(sum of deviations) for each of the $EE$ and $BB$ spectra and
show these in Fig.~\ref{fig:spectra}.
For the nine bandpowers used the expectation value/standard-deviation
for $\chi^2$ and $\chi$ are 9/4.2 and 0/3 respectively.
We see that BK150$\times$BK150 and BK150$\times$P353 are highly significant
in $BB$, while P353$\times$P353 has modest significance
in both $EE$ and $BB$.

\begin{figure}
\resizebox{0.95\columnwidth}{!}{\includegraphics{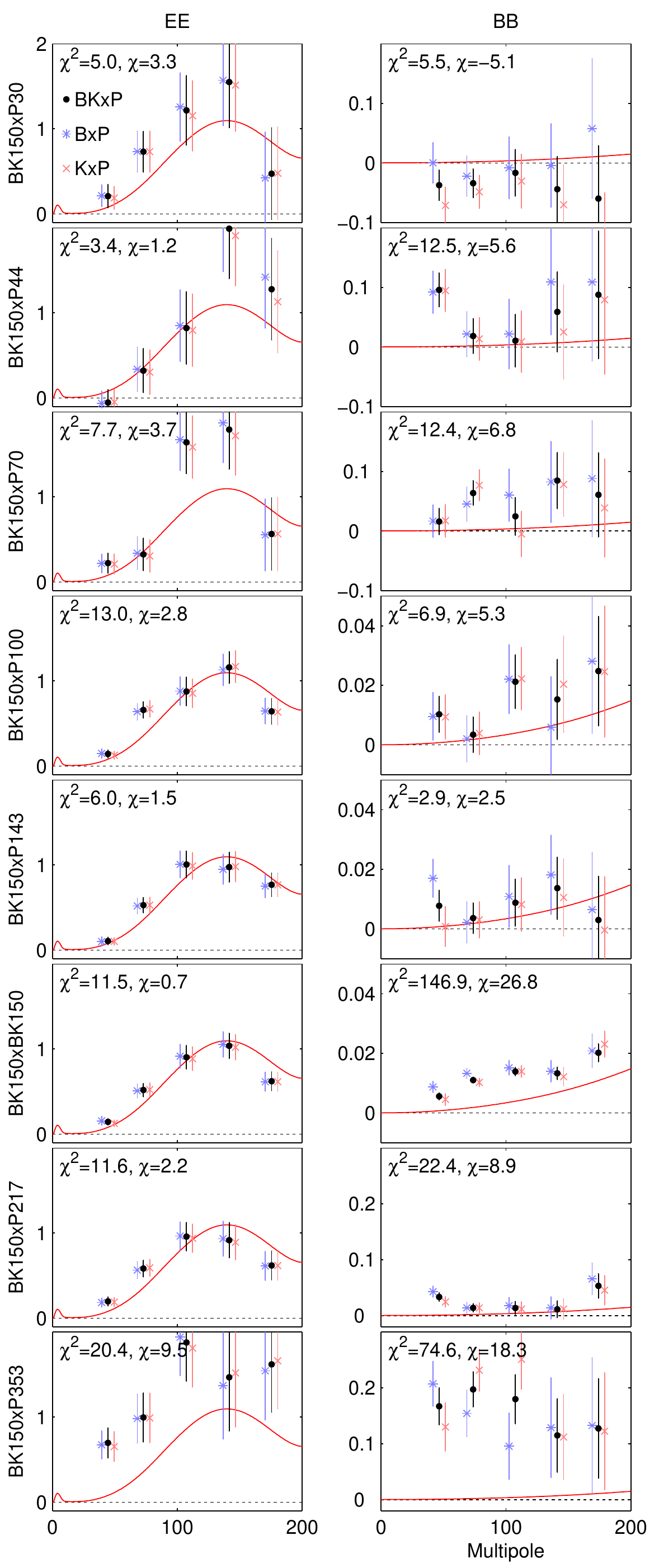}}
\caption{$EE$ (left column) and $BB$ (right column) cross-spectra between
\biceptwo/\keck\ maps and all of the polarized frequencies of \planck.
In all cases the quantity plotted is $\ell(\ell+1)C_l/2\pi$
in units of $\muK_{\mathrm{CMB}}^2$, and the red curves show the lensed-\lcdm\ expectations.
The error bars are the standard deviations of
lensed-\lcdm+noise simulations and hence contain no sample
variance on any other component.
Also note that the $y$-axis scales differ from panel to panel
in the right column.
The $\chi^2$ and $\chi$ (sum of deviations)
versus lensed-\lcdm\ for the five bandpowers shown
is marked at upper left.
There are no additional strong detections of
deviation from lensed-\lcdm\ over those
already shown in Fig.~\ref{fig:spectra}
although BK150$\times$P217 shows some evidence
of excess.}
\label{fig:spectra2}
\end{figure}

Figure~\ref{fig:spectra2} shows $EE$ and $BB$ cross-spectra between
\biceptwo/\keck\ and all of the polarized frequencies
of \planck\ (also including the \biceptwo/\keck\ auto-spectra).
For the five bandpowers shown the expectation value/standard-deviation
for $\chi^2$ and $\chi$ are 5/3.1 and 0/2.2 respectively.
As already noted, the BK150$\times$BK150 and BK150$\times$P353 $BB$ spectra show
highly significant excesses.
Additionally, there is evidence for excess $BB$ in
BK150$\times$P217 spectrum, and for excess $EE$ in BK150$\times$P353.
The other spectra in Fig.~\ref{fig:spectra2} show no strong evidence
for excess, although we note that only one of the $\chi$ values
is negative.
There is weak evidence for excess in the BK150$\times$P70 $BB$
spectrum but none in BK150$\times$P30 so this is presumably
just a noise fluctation.

There are a large number of additional \planck-only spectra, which are
not plotted here.
The noise on these is large and all are consistent with
\lcdm, with the possible exception of P217$\times$P353, where
modest evidence for an excess is seen in both $EE$ and $BB$ (see
e.g., Figure~10 of \piXXX ).

\subsection{Consistency of \biceptwo\ and \keckarray\ spectra}
\label{sec:speccons}

\begin{figure}
\resizebox{\columnwidth}{!}{\includegraphics{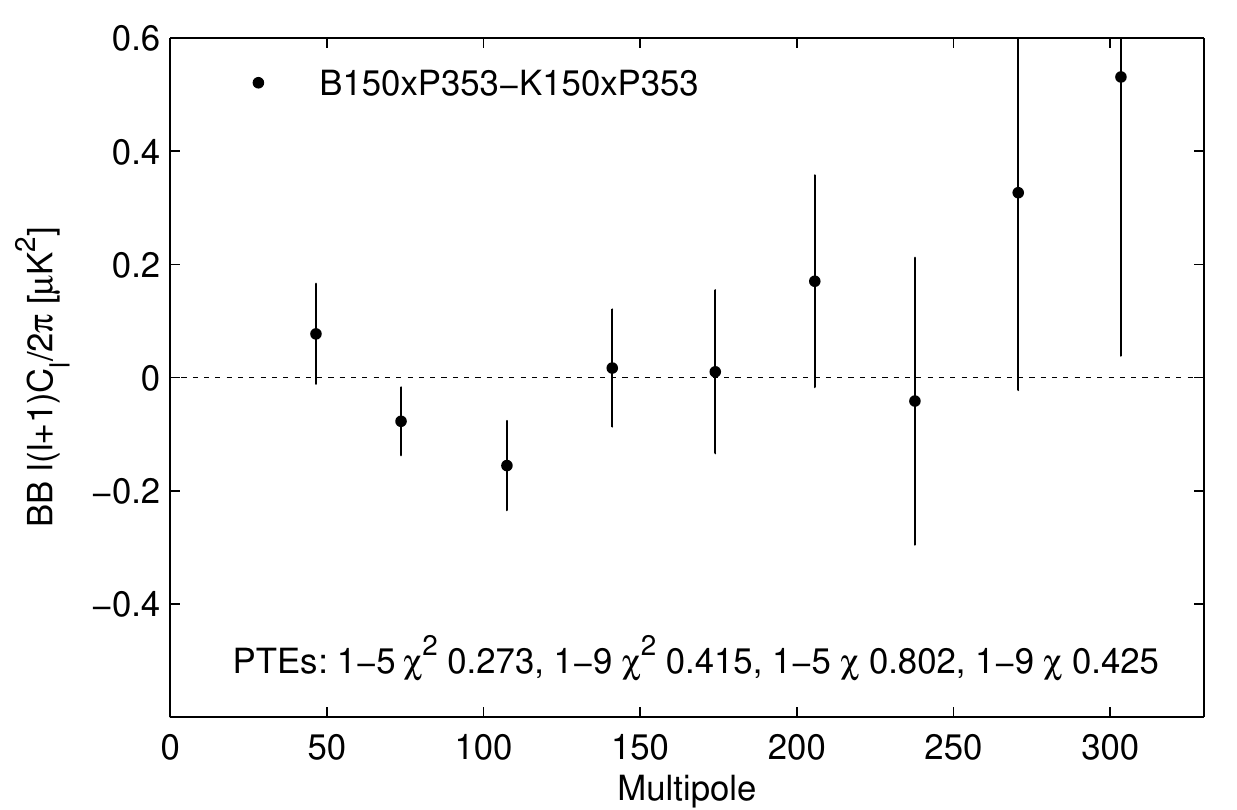}}
\caption{Differences of B150$\times$P353 and K150$\times$P353 $BB$ cross-spectra.
The error bars are the standard deviations of the pairwise
differences of signal+noise simulations that share common
input skies.
The probability to exceed the observed values of $\chi^2$
and $\chi$ statistics, as evaluated against the simulations, 
is quoted for bandpower ranges 1--5 ($20<\ell<200$) and 1--9 ($20<\ell<330$).
There is no evidence that these spectra are statistically incompatible.}
\label{fig:spectra3}
\end{figure}

The $BB$ auto-spectra for \biceptwo\ and \keckarray\
in the lower left panel of Fig.~\ref{fig:spectra} 
appear to differ by more than might be expected, given
that the \biceptwo\ and \keck\
maps cover almost exactly the same region of sky.
However, the error bars in this figure are the standard
deviations of lensed-\lcdm+noise simulations; while the signal is
largely common between the two experiments 
the noise is not, and the signal-noise cross terms produce
substantial additional fluctuation of the difference.
The correct way to quantify this is to compare the difference
of the real data to the pair-wise differences of simulations, 
using common input skies that have power similar to that observed
in the real data.
This was done in Section~8 of \bV\ and the \biceptwo\
and \keck\ maps were shown to be statistically compatible.
In an analogous manner we can also ask if the B150$\times$P353 and
K150$\times$P353 $BB$ cross-spectra shown in the bottom middle panel
of Fig.~\ref{fig:spectra} are compatible.
Figure~\ref{fig:spectra3} shows the results.
We calculate the $\chi^2$ and $\chi$ statistics on
these difference spectra and compare to the
simulated distributions exactly as in \bV.
The probability to exceed (PTE) the observed values 
is given in the figure for bandpowers 1--5 ($20<\ell<200$) and 1--9 ($20<\ell<330$).
There is no evidence that these spectra are statistically incompatible.

\subsection{Alternative power spectrum estimation}

\begin{figure}
\resizebox{\columnwidth}{!}{\includegraphics{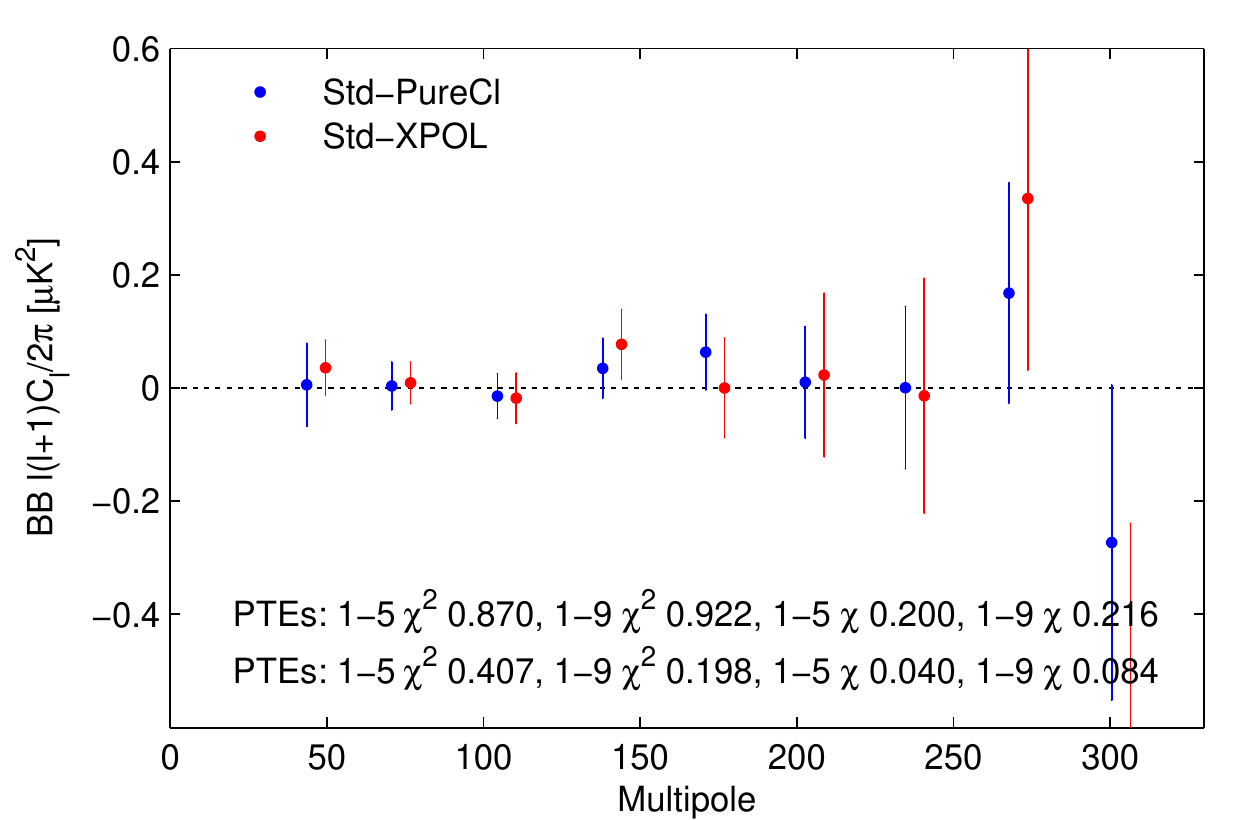}}
\caption{Differences of B150$\times$P353 $BB$ cross-spectra
from the standard power spectrum estimator and alternate
estimators.
The error bars are the standard deviations of the pairwise
differences of signal+noise simulations that share common
input skies.
The probability to exceed the observed values of $\chi^2$
and $\chi$ statistics, as evaluated against the simulations, 
is quoted for bandpower ranges 1--5 ($20<\ell<200$) and 1--9 ($20<\ell<330$).
We see that the differences of the real spectra are consistent
with the differences of the simulations.}
\label{fig:powspeccheck}
\end{figure}

We check the reliability of the power spectrum estimation with an
alternative pipeline.
The filtered and purified \planck\ and \biceptwo\ maps used
to make the spectra shown in Fig.~\ref{fig:spectra}
are transformed back into the \healpix\ pixelization using
cubic spline interpolation.  
The \bmode\ cross-power is then computed with the {\sc
  Xpol}~\cite{tristram2005} and {\sc PureCl}~\cite{Preece11}
estimators. 
Figure~\ref{fig:powspeccheck} shows the difference between these
alternative bandpowers and the standard bandpowers
for the B150$\times$P353 $BB$ cross spectrum.
As in Fig.~\ref{fig:spectra3} the errorbars are
the standard deviations of pairwise differences of simulations, 
which share common input skies and have power similar to that observed
in the real data.
The agreement is not expected to be exact due to the differing
bandpower window functions, but the differences of the real
bandpowers are consistent with those of the simulations.

\section{Likelihood analysis}
\label{sec:likeanal}

\subsection{Algorithm}
\label{sec:likemethod}

While it is conventional in plots like
Fig.~\ref{fig:spectra} to present bandpowers with
symmetric error bars, it is important to appreciate
that this is an approximation.
The likelihood of an observed bandpower
for a given model expectation value is generally an asymmetric
function, which can be computed given knowledge
of the noise level(s).
To compute the joint likelihood of an ensemble
of measured bandpower values it is of course
necessary to consider their full covariance---this
is especially important when using spectra
taken at different frequencies on the same field, 
where the signal covariance can be very strong.

We compute the bandpower covariance using
full simulations of signal-cross-signal,
noise-cross-noise, and signal-cross-noise.
From these, we can construct the covariance
matrix for a general model containing 
multiple signal components with any desired
set of SEDs.
When we do this we deliberately exclude terms
whose expectation value is zero, in order to reduce
noise in the resulting matrix due to the limited
number of simulated realizations.

To compute the joint likelihood of the data
for any given proposed model we use the
Hamimeche-Lewis~\cite{hamimeche08}
approximation (HL; see Section~9.1 of Ref.~\cite{barkats14} for
implementation details).
Here we extend the method to deal with single- and
cross-frequency spectra, and the covariances
thereof, in an analogous manner to the treatment
of, for example, $TT$, $TE$, and $EE$ in the standard
HL method.
The HL formulation requires that the bandpower covariance
matrix be determined for only a single ``fiducial model.''
We compute multi-dimensional grids of models explicitly
and/or use \texttt{COSMOMC} \citep{cosmomc} to sample the parameter space.

\subsection{Fiducial analysis}
\label{sec:likebase}

\begin{figure*}
\resizebox{0.9\textwidth}{!}{\includegraphics{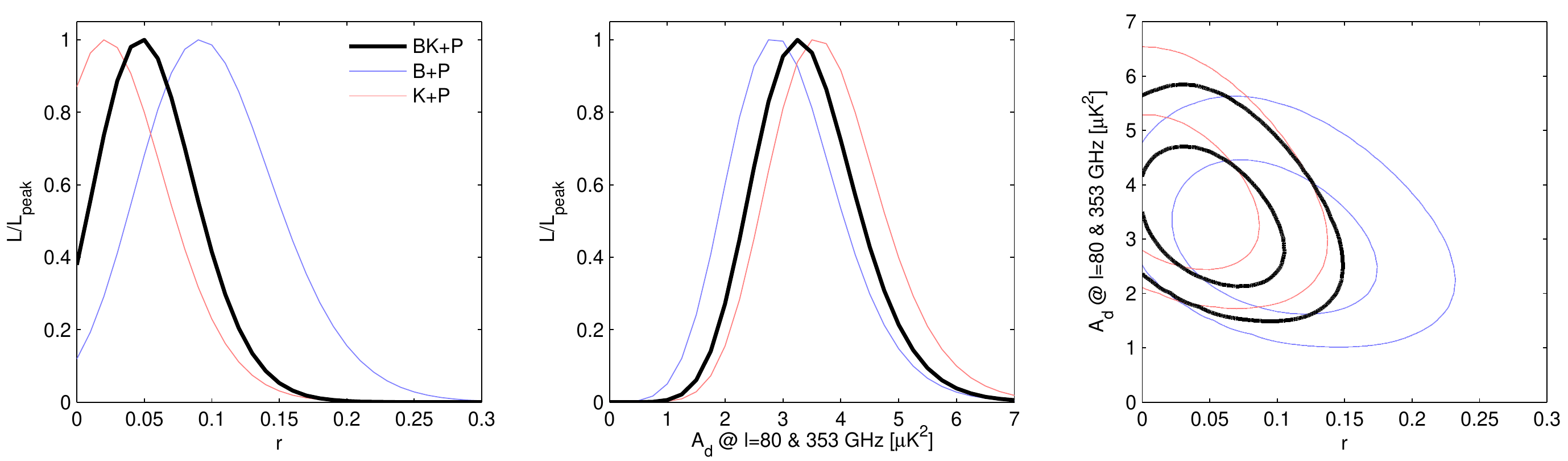}}
\caption{Likelihood results from a basic lensed-\lcdm+$r$+dust
model, fitting $BB$ auto- and cross-spectra taken between
maps at 150\,GHz, 217, and 353\,GHz.
The 217 and 353\,GHz maps come from \planck.
The primary results (heavy black) use the 150\,GHz
combined maps from \biceptwo/\keck.
Alternate curves (light blue and red) show how
the results vary when the \biceptwo\ and \keckarray\
only maps are used.
In all cases a Gaussian prior is placed on the dust frequency
spectrum parameter $\beta_\mathrm{d} = 1.59\pm0.11$.
In the right panel the two dimensional contours
enclose 68\% and 95\% of the total likelihood.}
\label{fig:likebase}
\end{figure*}

As an extension of the simplest lensed-\lcdm\ paradigm, we initially consider
a two component model of IGW with amplitude $r$, plus dust
with amplitude $A_\mathrm{d}$ (specified at 353\,GHz and $\ell=80$).
(Here we assume that the spectral index of the tensor modes ($n_\mathrm{t}$) is zero,
and a scalar pivot scale of $0.05\,\mathrm{Mpc}^{-1}$;
all values of $r$ quoted in this paper are $r_{0.05}$ unless noted
otherwise.)
Figure~\ref{fig:likebase} shows the results of fitting
such a model to $BB$ bandpowers taken between
\biceptwo/\keck\ and the 217 and 353\,GHz bands of \planck,
using bandpowers 1--5 ($20<\ell<200$).
For the \planck\ single-frequency case, the cross-spectrum of
detector-sets (DS1$\times$DS2) is used, following \piXXX.
The dust is modeled as a power law $\mathcal{D}_\ell \propto \ell^{-0.42}$,
with free amplitude $A_\mathrm{d}$ and scaling with frequency
according to the modified blackbody model.

As discussed in Sec.~\ref{sec:dustexpect} the simple
modified blackbody model is shown empirically in Ref.~\cite{planckiXXII}
to describe the mean polarized dust SED at mid Galactic latitudes
to an accuracy of a few percent over the frequency range 100--353\,GHz,
with variation of the $\beta_\mathrm{d}$ parameter being sufficient
to characterize the patch-to-patch variation.
Since it is not possible to constrain $\beta_\mathrm{d}$ using
the \biceptwo/\keck/ and \planck\ cross spectral bandpowers alone
a tight Gaussian prior $\beta_\mathrm{d} = 1.59\pm0.11$ is imposed,
the uncertainty being scaled from the observed patch-to-patch
variation at intermediate Galactic latitudes
in Ref.~\cite{planckiXXII}, as explained in \piXXX.
This prior assumes that the SED of dust polarization at intermediate 
latitudes~\cite{planckiXXII} applies to the high latitude \biceptwo/\keck\ field.
From dust astrophysics, we expect variations of the dust SED in intensity
and polarization to be correlated~\cite{martin07}.
We thus tested our assumption by measuring the $\beta_\mathrm{d}$ of the
dust total intensity in the \biceptwo/\keck\ field using
the template fitting analysis described in Ref.~\cite{planckiXVII},
and find the same value.

In Fig.~\ref{fig:likebase} we see that the \biceptwo\
data produce an $r$ likelihood that peaks 
higher than that for the \keckarray\ data.
This is because for $\ell<120$ the auto-spectrum B150$\times$B150 is 
higher than for K150$\times$K150, while the cross-spectrum B150$\times$P353
is lower than K150$\times$P353 (see Fig.~\ref{fig:spectra}).
However, recall that both pairs of spectra B150$\times$B150/K150$\times$K150 and
B150$\times$P353/K150$\times$P353 have been shown to be consistent within
noise fluctuation (see Sec.~\ref{sec:speccons}).
In Sec.~\ref{sec:validsim} these likelihood results
are also found to be compatible.
Given the consistency between the two experiments, the combined result
gives the best available measurement of the sky.

The combined curves (BK+P) in the left and center panels
of Fig.~\ref{fig:likebase} yield the following results:
$r=0.048^{+0.035}_{-0.032}$, $r<0.12$ at 95\% confidence,
and $A_\mathrm{d}=3.3^{+0.9}_{-0.8}$.
For $r$ the zero-to-peak likelihood ratio is 0.38.
Taking
$\frac{1}{2} \left( 1-f \left( -2\log{L_0/L_{\rm peak}} \right) \right)$,
where $f$ is the $\chi^2$ cdf (for one degree of freedom),
we estimate that the probability to get a number smaller than this is
8\% if in fact $r=0$. 
For $A_\mathrm{d}$ the zero-to-peak ratio is $1.8\times10^{-6}$ corresponding
to a smaller-than probability of $1.4\times10^{-7}$, and a $5.1\sigma$ detection
of dust power.

The maximum likelihood model on the grid has parameters $r=0.05$,
$A_\mathrm{d}=3.30$\,$\mu$K$^2$ (and $\beta_\mathrm{d}=1.6$).
Computing the bandpower covariance matrix for this
model, we obtain a $\chi^2$ of 40.8.
Using 28 degrees of freedom---5 bandpowers times
6 spectra, minus 2 fit parameters
(since $\beta_\mathrm{d}$ is not really free)---gives a PTE
of 0.06.
The largest contributions to $\chi^2$ come from the
P353$\times$P353 spectrum shown in the lower
right panel of Fig.~\ref{fig:spectra}.

\subsection{Variations from the fiducial data set and model}
\label{sec:likevar}

We now investigate a number of variations from the fiducial analysis
to see what difference these make to the constraint on $r$.

\begin{figure}
\resizebox{0.7\columnwidth}{!}{\includegraphics{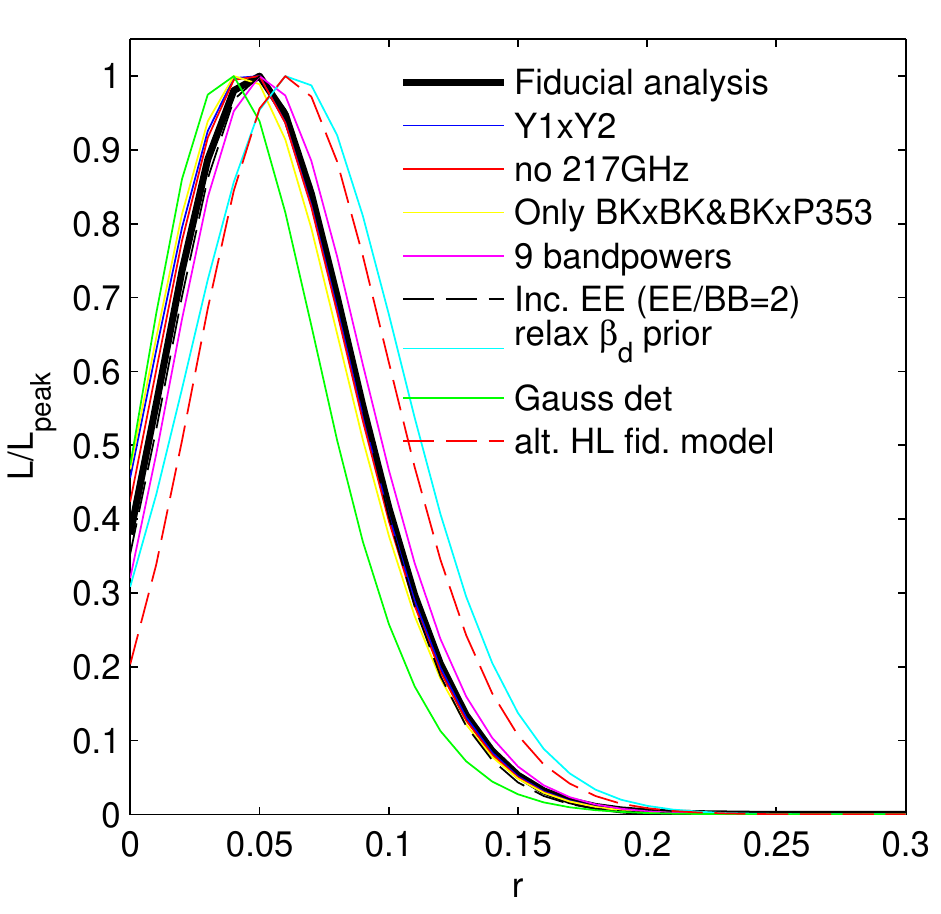}}
\caption{Likelihood results when varying the data sets
used and the model priors---see Sec~\ref{sec:likevar} for details.}
\label{fig:likevar}
\end{figure}

\begin{figure}
\resizebox{\columnwidth}{!}{\includegraphics{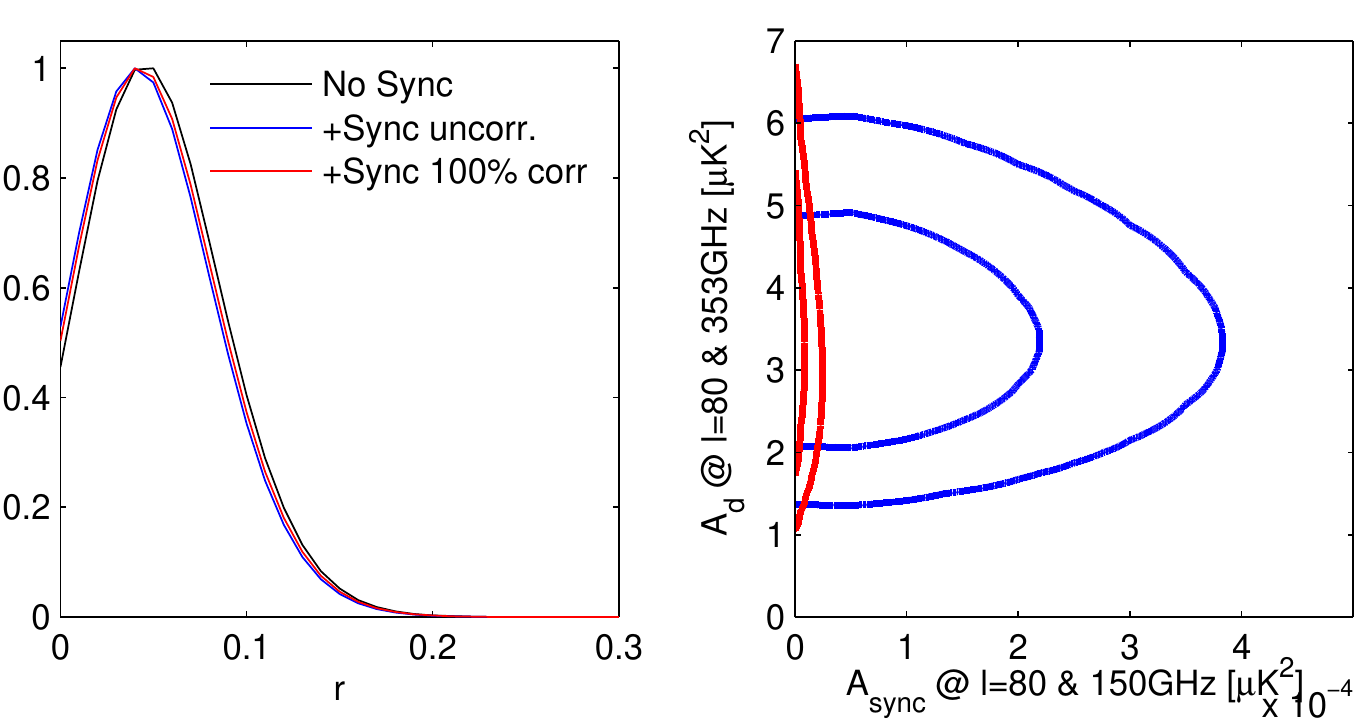}}
\caption{
Likelihood results for a fit when adding the lower frequency
bands of \planck, and extending the model to include a
synchrotron component.
The results for two different assumed degrees of
correlation between the dust and synchrotron sky patterns
are compared to those for the comparable model
without synchrotron (see text for details).}
\label{fig:sync}
\end{figure}

\begin{itemize}
\item {\bf Choice of \planck\ single-frequency spectra:}
switching the \planck\ single-frequency spectra to use one
of the alternative data splits (yearly or half-ring instead of detector-set) makes little
difference (see Fig.~\ref{fig:likevar}).
\item {\bf Using only 150 and 353\,GHz:}
dropping the spectra involving 217\,GHz from consideration
also has little effect (see Fig.~\ref{fig:likevar}).
\item{\bf Using only BK150$\times$BK150 and BK150$\times$P353:}
also excluding the 353\,GHz single-frequency spectrum
from consideration makes little difference.
The statistical weight of the BK150$\times$BK150 and BK150$\times$P353
spectra dominate (see Fig.~\ref{fig:likevar}).
\item {\bf Extending the bandpower range:}
going back to the base data set and 
extending the range of bandpowers considered to 1--9
(corresponding to $20<\ell<330$) makes
very little difference---the dominant statistical weight is with the
lower bandpowers (see Fig.~\ref{fig:likevar}).
\item{\bf Including $EE$ spectra:}
we can also include in the fits the $EE$ spectra shown in Fig.~\ref{fig:spectra2}.
\piXXX\ (figures~5 and~A.3) shows that the level of $EE$
from Galactic dust is on average around twice the level of $BB$.
However, there are substantial variations in this
ratio from sky-patch to sky-patch.
Setting $EE/BB=2$ we find that the
constraint on $A_\mathrm{d}$ narrows, while the $r$ constraint
changes little; this latter result is also shown in Fig.~\ref{fig:likevar}.
The maximum likelihood model on the grid is unchanged and
its $\chi^2$ PTE is acceptable.
\item {\bf Relaxing the $\beta_\mathrm{d}$ prior:}
relaxing the prior on the dust spectral index to
$\beta_\mathrm{d} = 1.59\pm0.33$ pushes
the peak of the $r$ constraint up 
(see Fig.~\ref{fig:likevar}).
However, it is not clear if this looser prior 
is self consistent; if the frequency spectral index varied
significantly across the sky it would invalidate cross-spectral
analysis, but there is strong evidence against 
such variation at high latitude, as explained in
Sec.~\ref{sec:dustdecorr}.
Nevertheless, it is important to appreciate that the $r$
constraint curves shown in Fig.~\ref{fig:likebase} shift
left (right) when assuming a lower (higher) value of
$\beta_\mathrm{d}$.
For $\beta_\mathrm{d} = 1.3\pm0.11$ the peak is at $r=0.021$
and for $\beta_\mathrm{d} = 1.9\pm0.11$  the peak is at $r=0.073$.
\item{\bf Varying the dust power spectrum shape:}
in the fiducial analysis the dust spatial power spectrum is
assumed to be a power law with $\mathcal{D}_\ell \propto \ell^{-0.42}$.
Marginalizing over spectral indices in the range $-0.8$ to 0 we find little
change in the $r$ constraint (
see also Sec.~\ref{sec:altanal} for an alternate
relaxation of the assumptions regarding the spatial
properties of the dust pattern).
\item {\bf Using Gaussian determinant likelihood:}
the fiducial analysis uses the HL likelihood
approximation, as described in Sec.~\ref{sec:likemethod}.
An alternative is to recompute the covariance matrix
$\mathbf{C}$ at each point in parameter space and take
$L=\det{( \mathbf{C})^{-1/2}} \exp{(-(\mathbf{d}^\mathrm{T} \mathbf{C}^{-1} \mathbf{d})/2)}$,
where $\mathbf{d}$ is the deviation of the observed bandpowers from the model
expectation values.
This results in an $r$ constraint which peaks 
slightly lower, as shown in Fig.~\ref{fig:likevar}.
Running both methods on the simulated realizations
described in Sec.~\ref{sec:validsim}, indicates that
such a difference is not unexpected and that
there may be a small systematic downward bias in the
Gaussian determinant method.
\item {\bf Varying the HL fiducial model:}
as mentioned in Sec.~\ref{sec:likemethod} the HL
likelihood formulation requires that the expectation
values and bandpower covariance matrix be provided for
a single ``fiducial model'' (not to be confused with
the ``fiducial analysis'' of Sec.~\ref{sec:likebase}).
The results from HL are supposed to be rather insensitive
to the choice of this model, although preferably
it should be close to reality.
Normally we use the lensed-\lcdm+dust simulations described
in Sec.~\ref{sec:validsim} below.
Switching this to lensed-\lcdm+$r$=0.2 produces no change
on average in the simulations, although it does
cause any given realization to shift slightly---the
change for the real data case is shown in
Fig.~\ref{fig:likevar}.
\item {\bf Adding synchrotron:}
\bI\ took the \wmap\ \textit{K}-band (23\,GHz) map,
extrapolated it to 150\,GHz according to $\nu^{-3.3}$
(mean value within the \biceptwo\ field of the MCMC
``Model f'' spectral index map provided by \wmap~\cite{bennett13}),
and found a negligible predicted contribution
($r_{\mathrm{sync,150}}=0.0008 \pm 0.0041$).
Figure~\ref{fig:spectra2} does not offer strong motivation
to reexamine this finding---the only
significant detections of correlated $BB$ power are
in the BK150$\times$P353 and, to a lesser extent, BK150$\times$P217 spectra.
However, here we proceed to a fit including all the
polarized bands of \planck\ (as shown in Fig.~\ref{fig:spectra2})
and adding a synchrotron
component to the base lensed-\lcdm+noise+$r$+dust model.
We take synchrotron to have a power law spectrum 
$\mathcal{D}_\ell \propto \ell^{-0.6}$~\cite{dunkley08}, 
with free amplitude $A_{\mathrm{sync}}$, where $A_{\mathrm{sync}}$ is
the amplitude at $\ell=80$ and at 150\,GHz, and scaling with frequency
according to $\nu^{-3.3}$. 
In such a scenario we can vary the degree of correlation
that is assumed between the dust and synchrotron sky
patterns.
Figure~\ref{fig:sync} shows results for the uncorrelated
and fully correlated cases.
Marginalizing over $r$ and $A_\mathrm{d}$ we find
$A_{\mathrm{sync}}<0.0003$\,$\mu$K$^2$ at 95\% confidence
for the uncorrelated case, and many times smaller for
the correlated.
This last is because once one has a detection of dust
it effectively becomes a template for the synchrotron.
This synchrotron limit is driven by the \planck\
30~GHz band---we obtain almost identical results when
adding only this band, and a much softer limit
when not including it.
If we instead assume synchrotron scaling of $\nu^{-3.0}$
the limit on $A_{\mathrm{sync}}$ is approximately
doubled for the uncorrelated case and reduced for the
correlated.
(Because the DS1$\times$DS2 data-split is not available
for the \planck\ LFI bands we switch to
Y1$\times$Y2 for this variant analysis, and so we compare
to this case in Fig.~\ref{fig:sync} rather than
the usual fiducial case.)
\item {\bf Varying lensing amplitude:}
in the fiducial analysis the amplitude of the lensing effect
is held fixed at the \lcdm\ expectation ($A_\mathrm{L}=1$).
Using their own and other data, the \planck\ Collaboration quote
a limit on the amplitude of the lensing effect versus
the \lcdm\ expectation of $A_\mathrm{L}=0.99\pm0.05$~\cite{planck2013XVI}.
Allowing $A_\mathrm{L}$ to float freely, and using all nine bandpowers, we
obtain the results shown in Fig.~\ref{fig:al}---there
is only weak degeneracy between $A_\mathrm{L}$ and both
$r$ and $A_\mathrm{d}$.
Marginalizing over $r$ and $A_\mathrm{d}$ we find
$A_\mathrm{L}=1.13\pm0.18$ with a likelihood ratio
between zero and peak of $3\times10^{-11}$.
Using the expression given in Sec.~\ref{sec:likebase} this corresponds
to a smaller-than probability of $2\times10^{-12}$,
equivalent to a $7.0\,\sigma$ detection of lensing in the $BB$
spectrum. 
We note this is the most significant to-date direct measurement of
lensing in \bmode\ polarization.
\end{itemize}

\begin{figure}
\resizebox{\columnwidth}{!}{\includegraphics{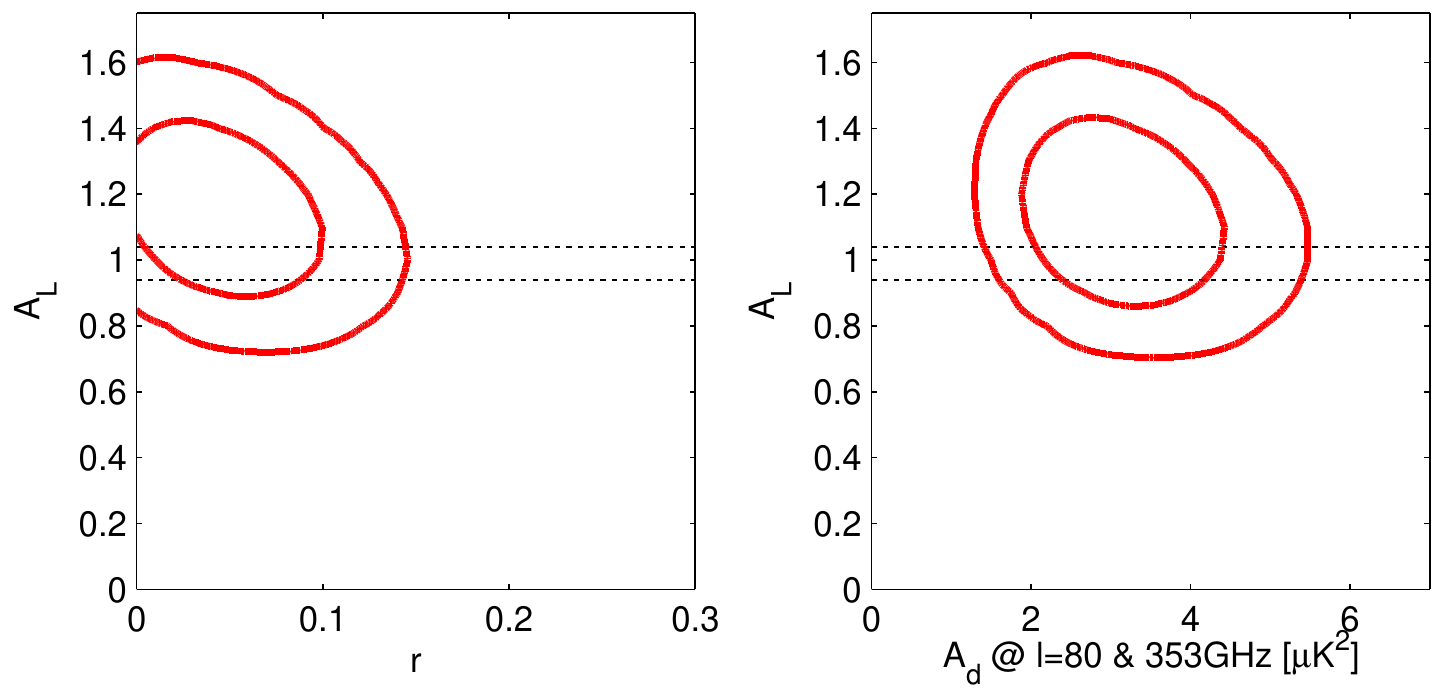}}
\caption{
Likelihood results for a fit allowing the lensing
scale factor $A_\mathrm{L}$ to float freely and using all nine bandpowers.
Marginalizing over $r$ and $A_\mathrm{d}$, we find that 
$A_\mathrm{L}=1.13\pm0.18$ and $A_\mathrm{L}=0$ is
ruled out with $7.0\,\sigma$ significance.}
\label{fig:al}
\end{figure}

\section{Likelihood validation}
\label{sec:valid}

\subsection{Validation with simulations}
\label{sec:validsim}

\begin{figure}
\resizebox{\columnwidth}{!}{\includegraphics{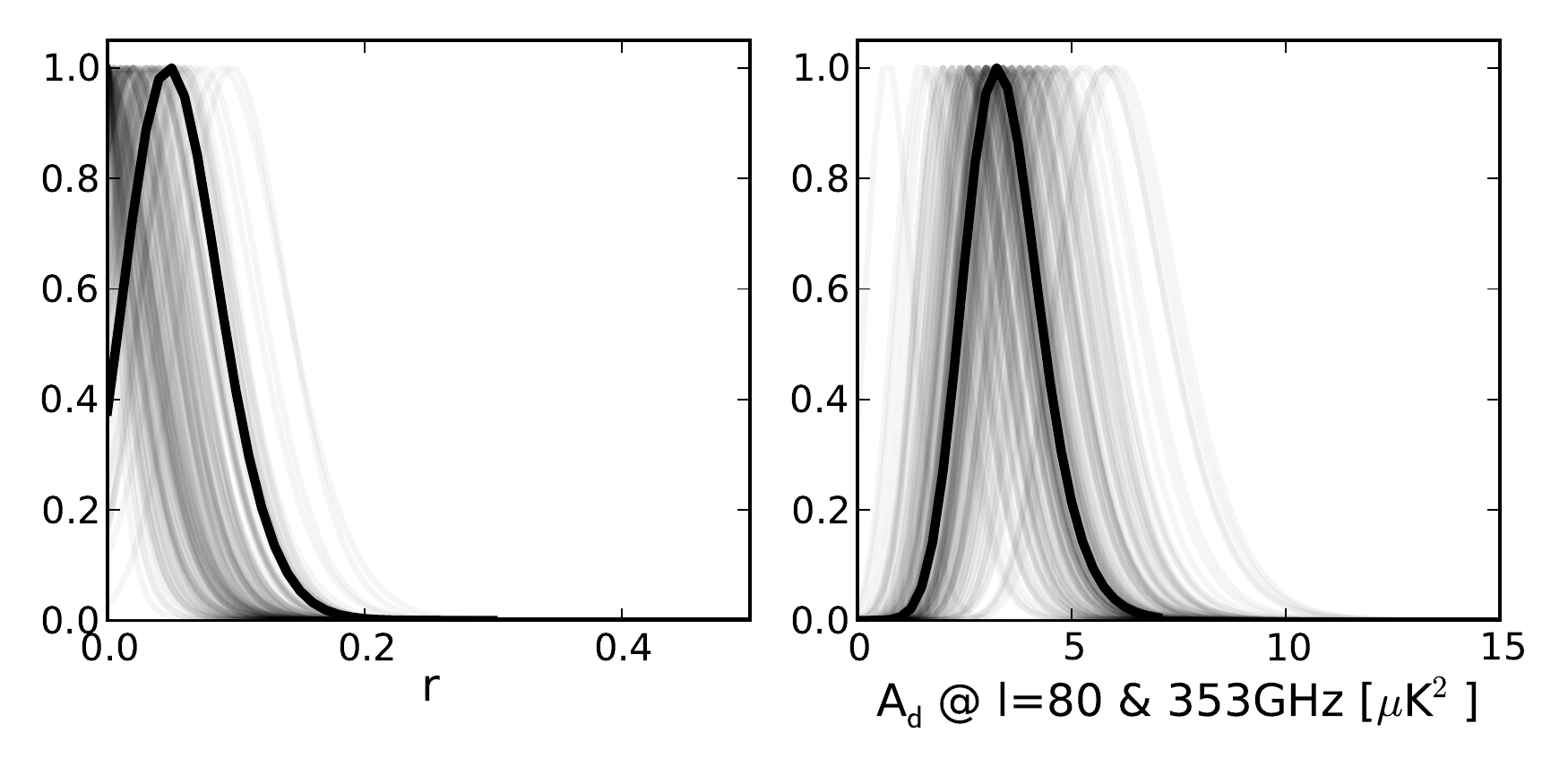}}
\caption{
Likelihoods for $r$ and $A_\mathrm{d}$, using \biceptwo/\keck\ and
\planck, 
as plotted in Fig.~\ref{fig:likebase}, overplotted on
constraints obtained from realizations of a
lensed-\lcdm+noise+dust model with dust power
similar to that favored by the real data
($A_\mathrm{d}=3.6$\,$\mu$K$^2$).
Half of the $r$ curves peak at zero as expected.}
\label{fig:simcons}
\end{figure}

We run the algorithm used in Sec.~\ref{sec:likebase}
on ensembles of simulated realizations to check its
performance.
We first consider a model where $r=0$ and $A_\mathrm{d}=3.6$\,$\mu$K$^2$,
this latter being close to the value favored by the data
in a dust-only scenario~\footnote{
Note that this is the number evaluated at 353\,GHz exactly---the
equivalent number as integrated over the \planck\ 353\,GHz passband
is 4.5\,$\mu$K$^2$ and the mask used in \piXXX\ is somewhat
different (larger) than the \biceptwo/\keck\ mask used here.}.
We generate Gaussian random realizations using
the fiducial spatial power law
$\mathcal{D}_\ell \propto \ell^{-0.42}$, scale these
to the various frequency bands using the modified
blackbody law with  $T_\mathrm{d}=19.6$\,K and $\beta_\mathrm{d} = 1.59$,
and add to the usual realizations of lensed-\lcdm+noise.
Figure~\ref{fig:simcons} shows some of the resulting $r$ and $A_\mathrm{d}$
constraint curves, 
with the result for the real data from Fig.~\ref{fig:likebase} overplotted.
As expected, approximately 50\% of the $r$ likelihoods peak above zero.
The median 95\% upper limit is $r<0.075$.
We find that $8$\% of the realizations have a ratio $L_0/L_{\rm peak}$ less
than the 0.38 observed in the real data, in agreement
with the estimate in Sec.~\ref{sec:likebase}.
Running these dust-only realizations for
\biceptwo\ only and \keckarray\ only, we find
that the shift in the maximum likelihood value of $r$
seen in the real data in Fig.~\ref{fig:likebase} is exceeded in
about 10\% of the simulations.

\begin{figure}
\resizebox{\columnwidth}{!}{\includegraphics{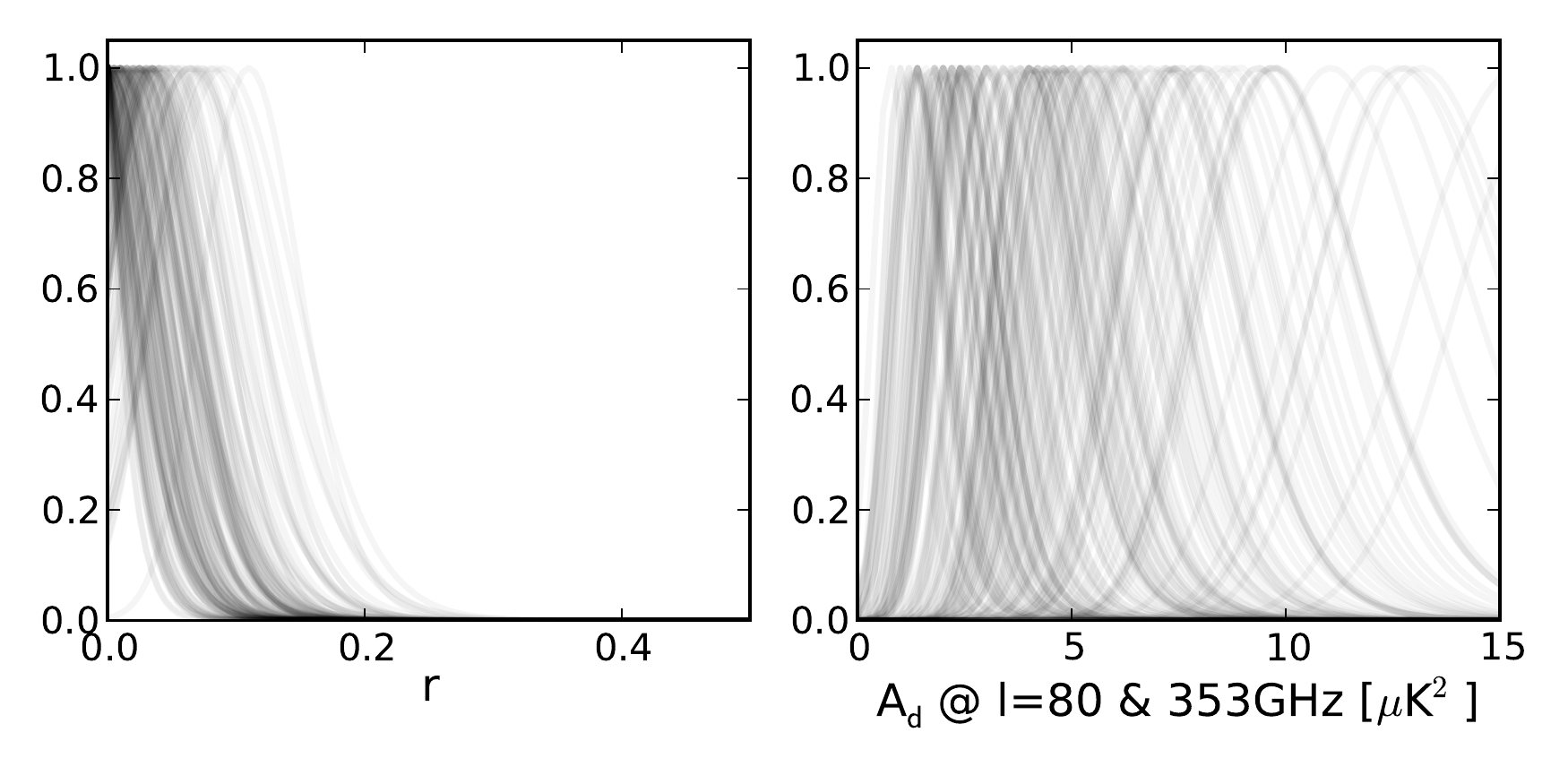}}
\caption{
Constraints obtained when adding dust realizations
from the \planck\ Sky Model version 1.7.8 to the base
lensed-\lcdm+noise simulations.
(Curves for 139 regions with peak $A_\mathrm{d} < 20$\,$\mu$K$^2$ are plotted.)
We see that the results for $r$ are unbiased in the presence
of dust realizations which do not necessarily follow the
$\ell^{-0.42}$ power law or have Gaussian fluctuations about it.}
\label{fig:psm}
\end{figure}

The above simulations assume that the dust component
follows on average the fiducial
$\mathcal{D}_\ell \propto \ell^{-0.42}$ spatial
power law, and fluctuates around it in a Gaussian manner.
To obtain sample dust sky patterns that may deviate from
this behavior in a way which better reflects reality,
we take the pre-launch version of the \planck\ Sky Model
(PSM; version 1.7.8 run in ``simulation'' mode)~\cite{delabrouille13}
evaluated in the \planck\ 353\,GHz band
and pull out the same 352 $|b|\,{>}\,35\deg$ partially
overlapping regions used in \piXXX.
We then scale these to the other bands and proceed as before.
Some of the regions have dust power orders of magnitude
higher than the real data and we cut them out
(selecting 139 regions with peak $A_\mathrm{d} < 20$\,$\mu$K$^2$).
Figure~\ref{fig:psm} presents the results.
The $r$ likelihoods will broaden as the level of $A_\mathrm{d}$
increases, and we should therefore not be surprised
if the fraction of realizations peaking at a value higher than
the real data is increased compared to the simulations
with mean $A_\mathrm{d}=3.6$\,$\mu$K$^2$.
However we still expect that on average 50\% will peak
above zero and approximately 8\% will have an $L_0/L_{\rm peak}$
ratio less than the 0.38 observed in the real data.
In fact we find 57\% and 7\%, respectively, consistent with
the expected values.
There is one realization which has a nominal
(false) detection of non-zero $r$ of $3.3\sigma$,
although this turns out to also have one of
the lowest $L_0/L_{\rm peak}$ ratios in the Gaussian
simulations shown in Fig.~\ref{fig:simcons}
(with which it shares the CMB and noise components),
so this is apparently just a relatively unlikely
fluctuation.

\subsection{Subtraction of scaled spectra}
\label{sec:altanal}

\begin{figure}
\resizebox{\columnwidth}{!}{\includegraphics{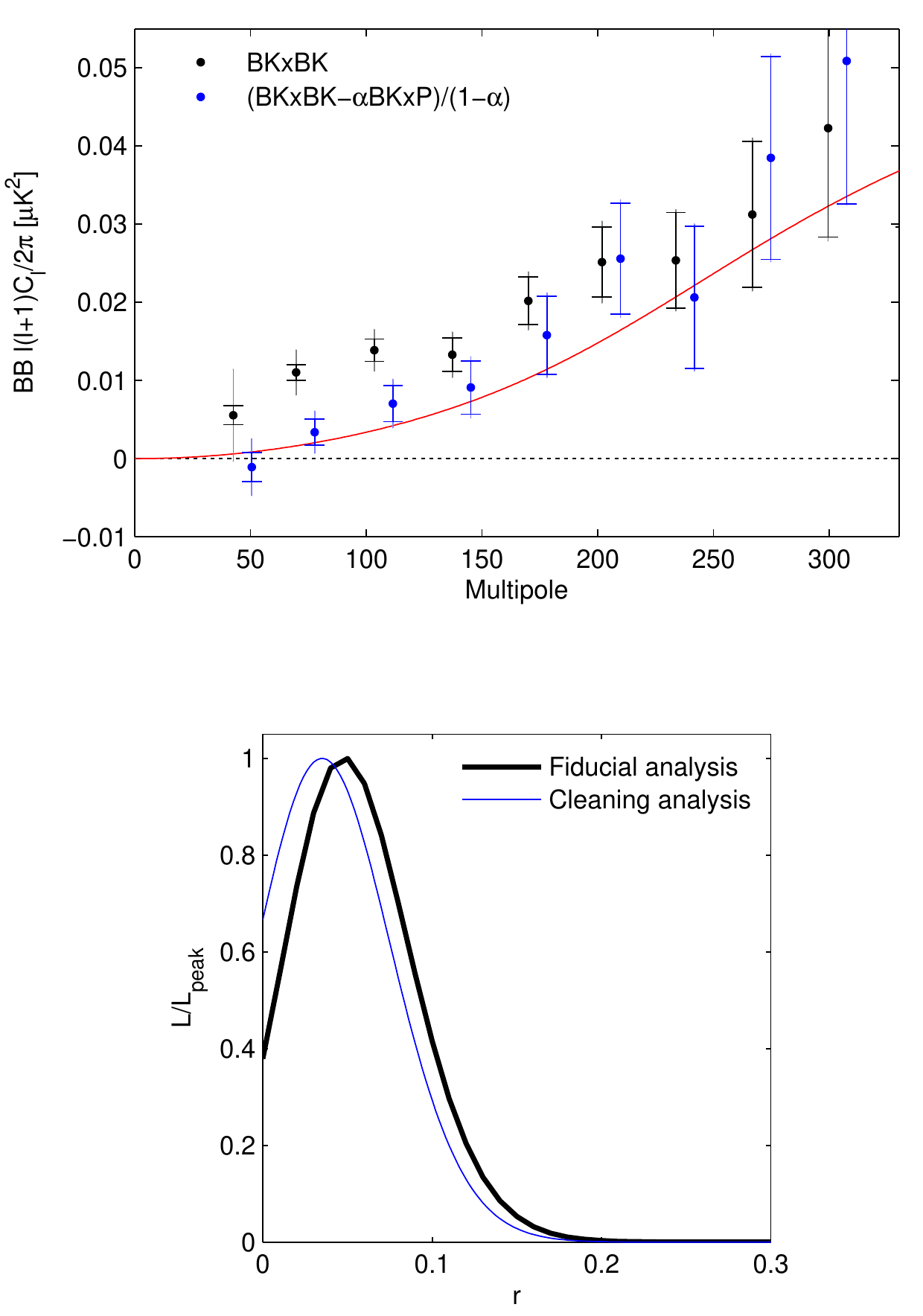}}
\caption{{\it Upper:} $BB$ spectrum of the \biceptwo/\keck\ maps
before and after subtraction of the dust contribution, estimated from
the cross-spectrum with \planck\ 353\,GHz.
The error bars are the standard deviations of
simulations, which, in the latter case, have been scaled and combined
in the same way.
The inner error bars are from lensed-\lcdm+noise simulations 
as in the previous plots, while the outer error bars are from the
lensed-\lcdm+noise+dust simulations.
The red curve shows the lensed-\lcdm\ expectation.
{\it Lower:} constraint on $r$ derived from the cleaned spectrum
compared to the fiducial analysis shown in Fig.~\ref{fig:likebase}.}
\label{fig:spectra4}
\end{figure}

As previously mentioned, the modified blackbody model predicts
that dust emission is 4\% as bright in the \biceptwo\ band
as it is in the \planck\ 353\,GHz band.
Therefore, taking the auto- and cross-spectra of the combined \biceptwo/\keck\ maps
and the \planck\ 353\,GHz maps, as shown in the bottom row of
Fig.~\ref{fig:spectra}, and evaluating 
$(\mathrm{BK}\times \mathrm{BK}-\alpha \mathrm{BK}\times \mathrm{P})/(1-\alpha)$,
at $\alpha=\alpha_\mathrm{fid}$ cleans out the dust contribution
(where $\alpha_\mathrm{fid}=0.04$).
The upper panel of Fig.~\ref{fig:spectra4} shows the result.

As an alternative to the full likelihood analysis presented
in Sec.~\ref{sec:likebase}, we can instead work with the
differenced spectra from above, a method we denote the ``cleaning'' approach.
If $\alpha_\mathrm{fid}$ were the true value, the expectation value of
this combination over CMB and noise would have no dust contribution.
However, dust would still contribute to its variance, but only
through its 2-point function.
In practice, we do not know $\alpha$ perfectly, and this  
uncertainty needs to be accounted for in a likelihood constructed from  
the differenced spectra.
Our approach is to treat the differenced spectra as a form of data
compression, and to compute the expectation value as a function of
$r$, $A_\mathrm{d}$, and $\beta_\mathrm{d}$ at each  
point in parameter space (the dust dependence enters for
$\alpha(\beta_\mathrm{d}) \neq \alpha_\mathrm{fid}$).
We use the method of Ref.~\cite{hamimeche08}, with a fiducial
covariance matrix, to build a likelihood for the difference spectra, 
and marginalize over $A_\mathrm{d}$ and
$\beta_\mathrm{d}$, and hence $\alpha$, adopting the prior
$\beta_\mathrm{d} = 1.59 \pm 0.11$.
This alternative likelihood has the advantage of being less sensitive to  
non-Gaussianity of the dust, since only the 2-point function of the  
dust affects the covariance of the differenced spectra close to
$\alpha_\mathrm{fid}$, while the full analysis may, in principle, be  
affected by the non-Gaussianity of the dust through 4-point  
contributions to power spectra covariances.
This cleaning approach does, however, ignore the (small amount of)
additional information available at other frequencies.
The lower panel of Fig.~\ref{fig:spectra4} compares the result to the
fiducial analysis with the full multi-spectra likelihood.
It is clear from the widths of the likelihood 
curves that compressing the spectra to form the cleaned difference  
results in very little loss of information on $r$.
The difference in peak values arises from the different data
treatments and is consistent with the scatter seen across simulations.
Finally, we note that one could also form a combination
$(\mathrm{BK}\times \mathrm{BK}-2\alpha \mathrm{BK}\times \mathrm{P} +
\alpha^2 \mathrm{P}\times \mathrm{P})/(1-\alpha)^2$
in which dust does not enter at all for
$\alpha=\alpha_\mathrm{fid}$.
However, the variance of this combination of  
spectra is large due to the \planck\ noise levels, and  
likelihoods built from this combination are considerably less  
constraining.

\section{Possible causes of decorrelation}
\label{sec:decorr}

Any systematic error that suppresses the BK150$\times$P353
cross-frequency spectrum with respect to the
BK150$\times$BK150 and P353$\times$P353 single-frequency
spectra would cause a systematic upward bias on the
$r$ constraint.
Here we investigate a couple of possibilities.

\subsection{Spatially varying dust frequency spectrum}
\label{sec:dustdecorr}

If the frequency dependence of polarized dust emission varied
from place to place on the sky, it would cause the
150\,GHz and 353\,GHz dust sky patterns to decorrelate
and suppress the BK150$\times$P353 cross-frequency spectrum
relative to the single-frequency spectra.
The assumption made so far in this paper is that such
decorrelation is negligible.
In fact \piXXX\ implicitly tests for such variation in their
Figure~6, where the \planck\ single- and cross-frequency spectra are
compared to the modified blackbody model (with the cross-frequency spectra
plotted at the geometric mean of their respective frequencies).
This plot is for an average over a large region of low foreground
sky (24\%); however, note that if there were spatial variation
of the spectral behavior anywhere in this region it would cause
suppression of the cross-frequency spectra with respect to
the single-frequency spectra.

\piXXX\ also tests explicitly for
evidence of decorrelation of the dust pattern across frequencies.
Their figure~E.1 shows the results for large and small sky patches.
The signal-to-noise ratio is low in clean regions, but no evidence
of decorrelation is found.

As a further check, we artificially suppress the amplitude of the
BK150$\times$P353 spectra in the Gaussian dust-only simulations
(see Sec.~\ref{sec:validsim}) by a conservative 10\%
(\piXXX\ sets a 7\% upper limit).
We find that the maximum likelihood value for $r$
shifts up by an average of 0.018, while
$A_\mathrm{d}$ shifts down by an average of 0.43\,$\mu$K$^2$, with
the size of the shift proportional to the magnitude of
the dust power in each given realization.
This behavior is readily understandable---since the
BK150$\times$BK150 and BK150$\times$P353 spectra dominate the
statistical weight, a decrease of the
latter is interpreted as a reduction in dust power, which
is compensated by an increase in $r$.
The bias on $r$ will be linearly related to the
assumed decorrelation factor.

\subsection{Calibration, analysis etc.}

Figure~\ref{fig:spectra2} shows that the $EE$ spectrum
BK150$\times$BK150 is extremely similar to that
for BK150$\times$P143.
We can compare such spectra to set limits on possible
decorrelation between the \biceptwo/\keck\ and \planck\
maps arising from any instrumental or analysis related effect, 
including differential pointing, polarization angle
mis-characterization, etc.
Taking the ratio of BK150$\times$P143 to the geometric mean of
BK150$\times$BK150 and P143H1$\times$P143H2, we find that for $TT$ the
decorrelation is approximately 0.1\%.
For $EE$ the signal-to-noise ratio is lower, but decorrelation
is limited to below 2\%, and consistent with zero when compared
to the fluctuation of signal+noise simulations.

\section{Conclusions}
\label{sec:conc}

\bI\ reported a highly significant detection of
\bmode\ polarization, at 150\,GHz, in excess of the
lensed-\lcdm\ expectation over the range $30<\ell<150$.
This excess has been confirmed by additional data
on the same field from the sucessor experiment
\keckarray.
\piXXX\ found that the level of dust
power in a field centered on the \biceptwo/\keck\ 
region (but somewhat larger than it) is of the
same magnitude as the reported excess,
but noted that, ``the present uncertainties are large,''
and that a joint analysis was required.

In this paper we have performed this joint
analysis, using the combined \biceptwo/\keck\ maps.
Cross-correlating these maps against all of the polarized
frequency bands of \planck\ we find a highly
significant \bmode\ detection only in the cross
spectrum with 353\,GHz.
We emphasize that this 150$\times$353\,GHz cross-spectrum has
a much higher signal-to-noise ratio than the 353\,GHz
single-frequency spectrum that \piXXX\ analyzed.

We have analyzed the data using a
multi-frequency, multi-component fit.
In this fit it is necessary to impose a
prior on the variation of the brightness of the
polarized dust emission with observing frequency, since the available
data are unable to constrain this alone, due to the relatively low
signal-to-noise ratio in \bmode\ polarization at 353\,GHz.
However based on the available information from \planck\
on the frequency dependence of polarized dust
emission across the mid and high Galactic latitude sky,
and the patch-to-patch stability thereof,
this prior appears to be justified and conservative.

We have shown that the final constraint on the
tensor-to-scalar ratio $r$ is very stable when
varying the frequency bands used, as well as the model priors.
The result does differ when using the \biceptwo\
and \keckarray\ data alone rather than in combination,
but the difference is compatible with noise fluctuation.
Expanding the model to include synchrotron emission,
while also including lower \planck\ frequencies, does
not change the result.

Allowing the amplitude of lensing to be free, we obtain
$A_\mathrm{L}=1.13 \pm 0.18$, with a significance of detection
of $7.0\sigma$. 
This is the most significant direct detection to-date of lensing in
\bmode\ polarization, even compared to experiments with higher angular
resolution.  
The \polarbear\ experiment has reported a
detection of \bmode\ lensing on smaller angular scales
($500<\ell<2100$), rejecting the $A_\mathrm{L}=0$ hypothesis at 97.2\% 
confidence~\cite{polarbear14}.
Additionally, \act~\cite{vanengelen14} and \spt~\cite{hanson13} have reported lensing
detections in polarization in cross-correlation with some other
tracer of the dark matter distribution on the sky.

We have validated the main likelihood
analysis on simulations of a dust-only model
and performed a simple subtraction of scaled
spectra, which approximates a map-based dust cleaning
(obtaining an $r$ constraint curve that peaks somewhat
lower).
Finally we investigated the possibility of astrophysical
or instrumental decorrelation of the sky patterns
between experiments or frequencies and find
no evidence for relevant bias.

The final result is expressed as a likelihood curve for $r$, and
yields an upper limit $r< 0.12$ at 95\% confidence.
The median limit in the lensed-\lcdm+noise+dust
simulations is $r<0.075$.
It is interesting to compare this latter to dust-free
simulations using only \biceptwo/\keck\ where the
median limit is $r<0.03$---the difference
represents the limitation due to noise in the \planck\ maps,
when marginalizing over dust.
The $r$ constraint curve peaks at  $r=0.05$ but disfavors
zero only by a factor of 2.5.
This is expected by chance 8\% of the time, as confirmed
in simulations of a dust-only model.
We emphasize that this significance is too
low to be interpreted as a detection of primordial \bmode s.
Transforming the \planck\ temperature-only 95\%
confidence limit of $r_{0.002}<0.11$~\cite{planck2013XVI}
to the pivot scale used in this paper
yields $r_{0.05}<0.12$, compatible with the present
result.

A \texttt{COSMOMC} module containing the bandpowers for
all cross-spectra between the combined \biceptwo/\keck\
maps and all of the frequencies of \planck\ is available
for download at \url{http://bicepkeck.org}.

\begin{figure}
\resizebox{\columnwidth}{!}{\includegraphics{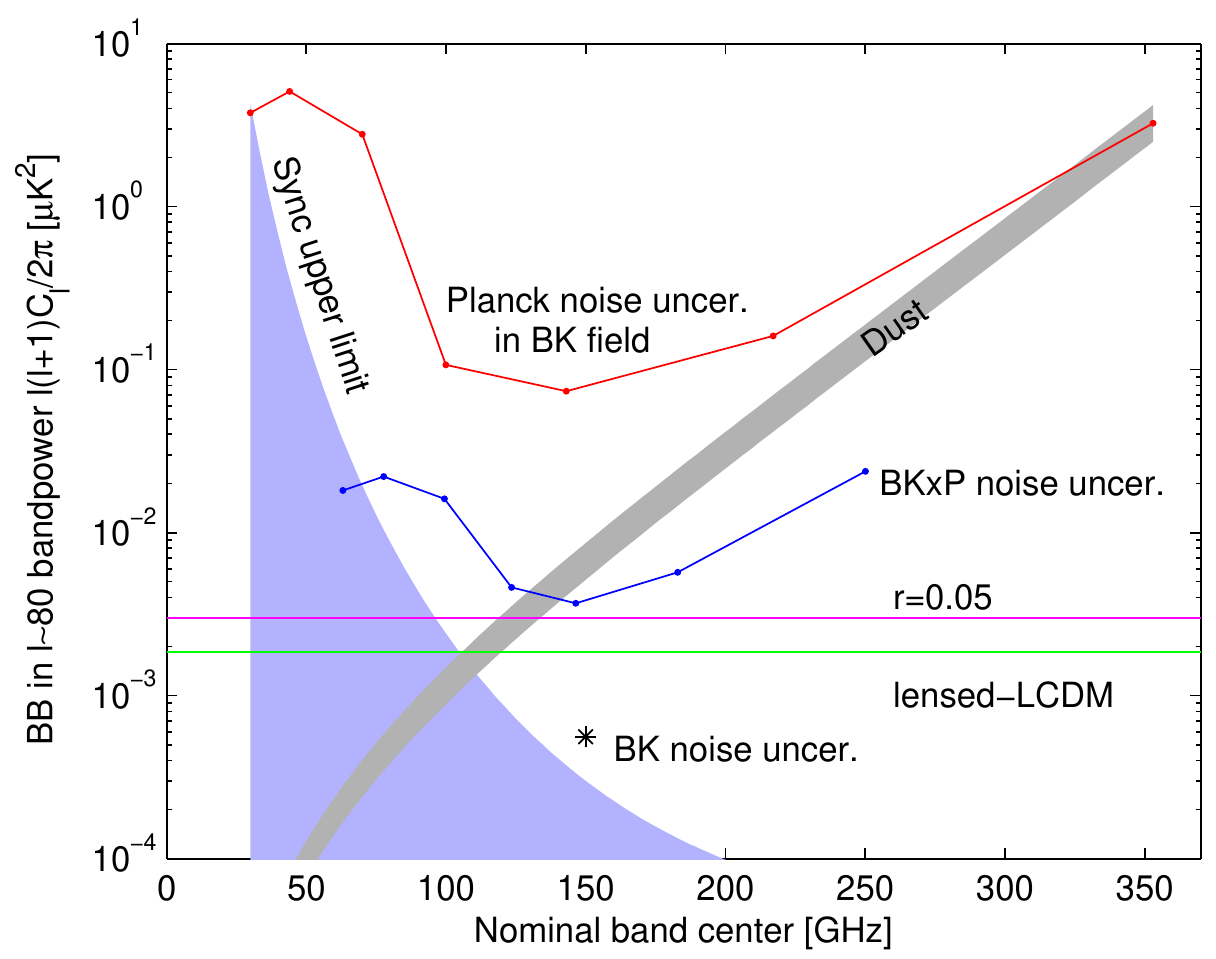}}
\caption{
Expectation values, and uncertainties thereon, for the $\ell\sim80$
$BB$  bandpower in the \biceptwo/\keck\ field.
The green and magenta lines correspond to the expected signal power of
lensed-\lcdm\ and $r=0.05$. 
Since CMB units are used, the levels corresponding
to these are flat with frequency.
The grey band shows the best fit
dust model (see Section~\ref{sec:likebase}) and
the blue shaded region shows the allowed region for synchrotron
(see Sec.~\ref{sec:likevar}).
The \biceptwo/\keck\ noise uncertainty is shown as a single starred
point, and the noise uncertainties of the \planck\ single-frequency
spectra evaluated in the \biceptwo/\keck\ field are shown in red.
The blue points show the noise uncertainty of the
cross-spectra taken between \biceptwo/\keck\ and,
from left to right, \planck\ 30, 44, 70, 100, 143, 217 \& 353~GHz,
and plotted at horizontal positions such that they
can be compared vertically with the dust and sync curves.}
\label{fig:noilev}
\end{figure}

In order to further constrain or detect IGW, additional data are 
required.
The \planck\ Collaboration may be able to make progress alone using the large
angular scale ``reionization bump,'' if systematics can be
appropriately controlled~\cite{efstathiou09}.
To take small patch ``recombination bump'' studies of the
type pursued here to the next level, data with
signal-to-noise comparable to that
achieved by \biceptwo/\keck\ at 150\,GHz are required at
more than one frequency.
Figure~\ref{fig:noilev} summarizes the situation.
The \biceptwo/\keck\ noise is much lower in the
\biceptwo/\keck\ field than the \planck\ noise.
However, since dust emission is dramatically brighter at
353\,GHz, it is detected in the cross-spectrum
between \biceptwo/\keck\ and \planck\ 353\,GHz.
Synchrotron is not detected and the crossover frequency with
dust is $\lesssim 100$\,GHz.
Planck's PR2 data release~\cite{planck2014-a12} shows that
for the cleanest 73\% of the sky, at 40' scales, the polarized foreground
minimum is at $\sim$80--90\,GHz.
During the 2014 season, two of the \keckarray\ receivers
observed in the 95\,GHz band and these data are under
active analysis.
\bicepthree\ will add substantial additional sensitivity
at 95\,GHz in the 2015, and especially 2016, seasons.
Meanwhile many other ground-based and sub-orbital experiments
are making measurements at a variety of frequencies
and sky coverage fractions.

\begin{acknowledgments}

\biceptwo\ was supported by the U.S. National Science Foundation under 
Grants No.\ ANT-0742818 and ANT-1044978 (Caltech and Harvard) 
and ANT-0742592 and ANT-1110087 (Chicago and Minnesota).  
The development of antenna-coupled detector technology was supported 
by the JPL Research and Technology Development Fund and Grants No.\
06-ARPA206-0040 and 10-SAT10-0017 from the NASA APRA and SAT programs.  
The \keckarray\ project was supported by the National Science
Foundation under Grants
ANT-1145172 and NSF-1255358 (Harvard), ANT-1145143 (Minnesota)
and ANT-1145248 (Stanford), and from the Keck Foundation
(Caltech).
We thank the staff of the U.S. Antarctic Program and in particular 
the South Pole Station without whose help this research would not have been possible.
The development of \planck\ has been supported by: ESA; CNES and
CNRS/INSU-IN2P3-INP (France); ASI, CNR, and INAF (Italy); NASA and DoE
(USA); STFC and UKSA (UK); CSIC, MICINN, JA and RES (Spain); Tekes,
AoF and CSC (Finland); DLR and MPG (Germany); CSA (Canada); DTU Space
(Denmark); SER/SSO (Switzerland); RCN (Norway); SFI (Ireland);
FCT/MCTES (Portugal); and ERC and PRACE (EU). A description of the Planck
Collaboration and a list of its members, including the technical or
scientific activities in which they have been involved, can be found
at
\url{http://www.sciops.esa.int/index.php?project=planck\&page=Planck_Collaboration.} 

\end{acknowledgments}

\bibliography{ms}

\begin{thebibliography}{50}%
\makeatletter
\providecommand \@ifxundefined [1]{%
 \@ifx{#1\undefined}
}%
\providecommand \@ifnum [1]{%
 \ifnum #1\expandafter \@firstoftwo
 \else \expandafter \@secondoftwo
 \fi
}%
\providecommand \@ifx [1]{%
 \ifx #1\expandafter \@firstoftwo
 \else \expandafter \@secondoftwo
 \fi
}%
\providecommand \natexlab [1]{#1}%
\providecommand \enquote  [1]{``#1''}%
\providecommand \bibnamefont  [1]{#1}%
\providecommand \bibfnamefont [1]{#1}%
\providecommand \citenamefont [1]{#1}%
\providecommand \href@noop [0]{\@secondoftwo}%
\providecommand \href [0]{\begingroup \@sanitize@url \@href}%
\providecommand \@href[1]{\@@startlink{#1}\@@href}%
\providecommand \@@href[1]{\endgroup#1\@@endlink}%
\providecommand \@sanitize@url [0]{\catcode `\\12\catcode `\$12\catcode
  `\&12\catcode `\#12\catcode `\^12\catcode `\_12\catcode `\%12\relax}%
\providecommand \@@startlink[1]{}%
\providecommand \@@endlink[0]{}%
\providecommand \url  [0]{\begingroup\@sanitize@url \@url }%
\providecommand \@url [1]{\endgroup\@href {#1}{\urlprefix }}%
\providecommand \urlprefix  [0]{URL }%
\providecommand \Eprint [0]{\href }%
\providecommand \doibase [0]{http://dx.doi.org/}%
\providecommand \selectlanguage [0]{\@gobble}%
\providecommand \bibinfo  [0]{\@secondoftwo}%
\providecommand \bibfield  [0]{\@secondoftwo}%
\providecommand \translation [1]{[#1]}%
\providecommand \BibitemOpen [0]{}%
\providecommand \bibitemStop [0]{}%
\providecommand \bibitemNoStop [0]{.\EOS\space}%
\providecommand \EOS [0]{\spacefactor3000\relax}%
\providecommand \BibitemShut  [1]{\csname bibitem#1\endcsname}%
\let\auto@bib@innerbib\@empty
\bibitem [{\citenamefont {{Penzias}}\ and\ \citenamefont
  {{Wilson}}(1965)}]{penzias65}%
  \BibitemOpen
  \bibfield  {author} {\bibinfo {author} {\bibfnamefont {A.~A.}\ \bibnamefont
  {{Penzias}}}\ and\ \bibinfo {author} {\bibfnamefont {R.~W.}\ \bibnamefont
  {{Wilson}}},\ }\href {\doibase 10.1086/148307} {\bibfield  {journal}
  {\bibinfo  {journal} {\apj}\ }\textbf {\bibinfo {volume} {142}},\ \bibinfo
  {pages} {419} (\bibinfo {year} {1965})}\BibitemShut {NoStop}%
\bibitem [{\citenamefont {{Bennett}}\ \emph {et~al.}(2013)\citenamefont
  {{Bennett}}, \citenamefont {{Larson}}, \citenamefont {{Weiland}},
  \citenamefont {{Jarosik}}, \citenamefont {{Hinshaw}}, \citenamefont
  {{Odegard}}, \citenamefont {{Smith}}, \citenamefont {{Hill}}, \citenamefont
  {{Gold}}, \citenamefont {{Halpern}}, \citenamefont {{Komatsu}}, \citenamefont
  {{Nolta}}, \citenamefont {{Page}}, \citenamefont {{Spergel}}, \citenamefont
  {{Wollack}}, \citenamefont {{Dunkley}}, \citenamefont {{Kogut}},
  \citenamefont {{Limon}}, \citenamefont {{Meyer}}, \citenamefont {{Tucker}},\
  and\ \citenamefont {{Wright}}}]{bennett13}%
  \BibitemOpen
  \bibfield  {author} {\bibinfo {author} {\bibfnamefont {C.~L.}\ \bibnamefont
  {{Bennett}}}, \bibinfo {author} {\bibfnamefont {D.}~\bibnamefont {{Larson}}},
  \bibinfo {author} {\bibfnamefont {J.~L.}\ \bibnamefont {{Weiland}}}, \bibinfo
  {author} {\bibfnamefont {N.}~\bibnamefont {{Jarosik}}}, \bibinfo {author}
  {\bibfnamefont {G.}~\bibnamefont {{Hinshaw}}}, \bibinfo {author}
  {\bibfnamefont {N.}~\bibnamefont {{Odegard}}}, \bibinfo {author}
  {\bibfnamefont {K.~M.}\ \bibnamefont {{Smith}}}, \bibinfo {author}
  {\bibfnamefont {R.~S.}\ \bibnamefont {{Hill}}}, \bibinfo {author}
  {\bibfnamefont {B.}~\bibnamefont {{Gold}}}, \bibinfo {author} {\bibfnamefont
  {M.}~\bibnamefont {{Halpern}}}, \bibinfo {author} {\bibfnamefont
  {E.}~\bibnamefont {{Komatsu}}}, \bibinfo {author} {\bibfnamefont {M.~R.}\
  \bibnamefont {{Nolta}}}, \bibinfo {author} {\bibfnamefont {L.}~\bibnamefont
  {{Page}}}, \bibinfo {author} {\bibfnamefont {D.~N.}\ \bibnamefont
  {{Spergel}}}, \bibinfo {author} {\bibfnamefont {E.}~\bibnamefont
  {{Wollack}}}, \bibinfo {author} {\bibfnamefont {J.}~\bibnamefont
  {{Dunkley}}}, \bibinfo {author} {\bibfnamefont {A.}~\bibnamefont {{Kogut}}},
  \bibinfo {author} {\bibfnamefont {M.}~\bibnamefont {{Limon}}}, \bibinfo
  {author} {\bibfnamefont {S.~S.}\ \bibnamefont {{Meyer}}}, \bibinfo {author}
  {\bibfnamefont {G.~S.}\ \bibnamefont {{Tucker}}}, \ and\ \bibinfo {author}
  {\bibfnamefont {E.~L.}\ \bibnamefont {{Wright}}},\ }\href {\doibase
  10.1088/0067-0049/208/2/20} {\bibfield  {journal} {\bibinfo  {journal}
  {\apjs}\ }\textbf {\bibinfo {volume} {208}},\ \bibinfo {eid} {20} (\bibinfo
  {year} {2013})},\ \Eprint {http://arxiv.org/abs/1212.5225} {arXiv:1212.5225}
  \BibitemShut {NoStop}%
\bibitem [{\citenamefont {{Planck Collaboration XVI}}(2014)}]{planck2013XVI}%
  \BibitemOpen
  \bibfield  {author} {\bibinfo {author} {\bibnamefont {{Planck Collaboration
  XVI}}},\ }\href {\doibase 10.1051/0004-6361/201321591} {\bibfield  {journal}
  {\bibinfo  {journal} {\aap}\ }\textbf {\bibinfo {volume} {571}},\ \bibinfo
  {eid} {A16} (\bibinfo {year} {2014})},\ \Eprint
  {http://arxiv.org/abs/1303.5076} {arXiv:1303.5076 [astro-ph.CO]} \BibitemShut
  {NoStop}%
\bibitem [{\citenamefont {{Planck Collaboration XXII}}(2014)}]{planck2013XXII}%
  \BibitemOpen
  \bibfield  {author} {\bibinfo {author} {\bibnamefont {{Planck Collaboration
  XXII}}},\ }\href@noop {} {\bibfield  {journal} {\bibinfo  {journal} {\aap}\
  }\textbf {\bibinfo {volume} {571}},\ \bibinfo {eid} {A22} (\bibinfo {year}
  {2014})},\ \Eprint {http://arxiv.org/abs/1303.5082} {arXiv:1303.5082}
  \BibitemShut {NoStop}%
\bibitem [{\citenamefont {{Starobinsky}}(1979)}]{starobinsky79}%
  \BibitemOpen
  \bibfield  {author} {\bibinfo {author} {\bibfnamefont {A.~A.}\ \bibnamefont
  {{Starobinsky}}},\ }\href@noop {} {\bibfield  {journal} {\bibinfo  {journal}
  {Zh.\ Eksp.\ Teor.\ Fiz.\ Pisma Red.}\ }\textbf {\bibinfo {volume} {30}},\
  \bibinfo {pages} {719} (\bibinfo {year} {1979})}\BibitemShut {NoStop}%
\bibitem [{\citenamefont {{Rubakov}}\ \emph {et~al.}(1982)\citenamefont
  {{Rubakov}}, \citenamefont {{Sazhin}},\ and\ \citenamefont
  {{Veryaskin}}}]{rubakov82}%
  \BibitemOpen
  \bibfield  {author} {\bibinfo {author} {\bibfnamefont {V.~A.}\ \bibnamefont
  {{Rubakov}}}, \bibinfo {author} {\bibfnamefont {M.~V.}\ \bibnamefont
  {{Sazhin}}}, \ and\ \bibinfo {author} {\bibfnamefont {A.~V.}\ \bibnamefont
  {{Veryaskin}}},\ }\href {\doibase 10.1016/0370-2693(82)90641-4} {\bibfield
  {journal} {\bibinfo  {journal} {Phys.\ Lett.\ B}\ }\textbf {\bibinfo {volume}
  {115}},\ \bibinfo {pages} {189} (\bibinfo {year} {1982})}\BibitemShut
  {NoStop}%
\bibitem [{\citenamefont {{Fabbri}}\ and\ \citenamefont
  {{Pollock}}(1983)}]{fabbri83}%
  \BibitemOpen
  \bibfield  {author} {\bibinfo {author} {\bibfnamefont {R.}~\bibnamefont
  {{Fabbri}}}\ and\ \bibinfo {author} {\bibfnamefont {M.~D.}\ \bibnamefont
  {{Pollock}}},\ }\href {\doibase 10.1016/0370-2693(83)91322-9} {\bibfield
  {journal} {\bibinfo  {journal} {Phys.\ Lett.\ B}\ }\textbf {\bibinfo {volume}
  {125}},\ \bibinfo {pages} {445} (\bibinfo {year} {1983})}\BibitemShut
  {NoStop}%
\bibitem [{\citenamefont {{Abbott}}\ and\ \citenamefont
  {{Wise}}(1984)}]{abbott84}%
  \BibitemOpen
  \bibfield  {author} {\bibinfo {author} {\bibfnamefont {L.~F.}\ \bibnamefont
  {{Abbott}}}\ and\ \bibinfo {author} {\bibfnamefont {M.~B.}\ \bibnamefont
  {{Wise}}},\ }\href {\doibase 10.1016/0550-3213(84)90329-8} {\bibfield
  {journal} {\bibinfo  {journal} {Nucl.\ Phys.\ B}\ }\textbf {\bibinfo {volume}
  {244}},\ \bibinfo {pages} {541} (\bibinfo {year} {1984})}\BibitemShut
  {NoStop}%
\bibitem [{\citenamefont {{Polnarev}}(1985)}]{polnarev85}%
  \BibitemOpen
  \bibfield  {author} {\bibinfo {author} {\bibfnamefont {A.~G.}\ \bibnamefont
  {{Polnarev}}},\ }\href@noop {} {\bibfield  {journal} {\bibinfo  {journal}
  {\sovast}\ }\textbf {\bibinfo {volume} {29}},\ \bibinfo {pages} {607}
  (\bibinfo {year} {1985})}\BibitemShut {NoStop}%
\bibitem [{\citenamefont {{Seljak}}\ and\ \citenamefont
  {{Zaldarriaga}}(1997)}]{seljak97a}%
  \BibitemOpen
  \bibfield  {author} {\bibinfo {author} {\bibfnamefont {U.}~\bibnamefont
  {{Seljak}}}\ and\ \bibinfo {author} {\bibfnamefont {M.}~\bibnamefont
  {{Zaldarriaga}}},\ }\href {\doibase 10.1103/PhysRevLett.78.2054} {\bibfield
  {journal} {\bibinfo  {journal} {\prl}\ }\textbf {\bibinfo {volume} {78}},\
  \bibinfo {pages} {2054} (\bibinfo {year} {1997})},\ \Eprint
  {http://arxiv.org/abs/astro-ph/9609169} {astro-ph/9609169} \BibitemShut
  {NoStop}%
\bibitem [{\citenamefont {{Seljak}}(1997)}]{seljak97b}%
  \BibitemOpen
  \bibfield  {author} {\bibinfo {author} {\bibfnamefont {U.}~\bibnamefont
  {{Seljak}}},\ }\href {\doibase 10.1086/304123} {\bibfield  {journal}
  {\bibinfo  {journal} {\apj}\ }\textbf {\bibinfo {volume} {482}},\ \bibinfo
  {pages} {6} (\bibinfo {year} {1997})},\ \Eprint
  {http://arxiv.org/abs/astro-ph/9608131} {astro-ph/9608131} \BibitemShut
  {NoStop}%
\bibitem [{\citenamefont {{Kamionkowski}}\ \emph {et~al.}(1997)\citenamefont
  {{Kamionkowski}}, \citenamefont {{Kosowsky}},\ and\ \citenamefont
  {{Stebbins}}}]{kamionkowski97}%
  \BibitemOpen
  \bibfield  {author} {\bibinfo {author} {\bibfnamefont {M.}~\bibnamefont
  {{Kamionkowski}}}, \bibinfo {author} {\bibfnamefont {A.}~\bibnamefont
  {{Kosowsky}}}, \ and\ \bibinfo {author} {\bibfnamefont {A.}~\bibnamefont
  {{Stebbins}}},\ }\href {\doibase 10.1103/PhysRevLett.78.2058} {\bibfield
  {journal} {\bibinfo  {journal} {\prl}\ }\textbf {\bibinfo {volume} {78}},\
  \bibinfo {pages} {2058} (\bibinfo {year} {1997})},\ \Eprint
  {http://arxiv.org/abs/astro-ph/9609132} {astro-ph/9609132} \BibitemShut
  {NoStop}%
\bibitem [{Note2()}]{Note2}%
  \BibitemOpen
  \bibinfo {note} {{\protect \it Planck}\ (\protect \url
  {http://www.esa.int/planck}) is a project of the European Space Agency (ESA)
  with instruments provided by two scientific consortia funded by ESA member
  states (in particular the lead countries, France and Italy) with
  contributions from NASA (USA), and telescope reflectors provided in a
  collaboration between ESA and a scientific consortium led and funded by
  Denmark.}\BibitemShut {Stop}%
\bibitem [{\citenamefont {{Planck Collaboration 2011A}}(2011)}]{planck2011I}%
  \BibitemOpen
  \bibfield  {author} {\bibinfo {author} {\bibnamefont {{Planck Collaboration
  2011A}}},\ }\href@noop {} {\bibfield  {journal} {\bibinfo  {journal} {\aap}\
  }\textbf {\bibinfo {volume} {536}},\ \bibinfo {pages} {A1} (\bibinfo {year}
  {2011})}\BibitemShut {NoStop}%
\bibitem [{\citenamefont {{Planck Collaboration I}}(2014)}]{planck2013I}%
  \BibitemOpen
  \bibfield  {author} {\bibinfo {author} {\bibnamefont {{Planck Collaboration
  I}}},\ }\href@noop {} {\bibfield  {journal} {\bibinfo  {journal} {\aap}\
  }\textbf {\bibinfo {volume} {571}},\ \bibinfo {pages} {A1} (\bibinfo {year}
  {2014})}\BibitemShut {NoStop}%
\bibitem [{\citenamefont {{Hildebrand}}\ \emph {et~al.}(1999)\citenamefont
  {{Hildebrand}}, \citenamefont {{Dotson}}, \citenamefont {{Dowell}},
  \citenamefont {{Schleuning}},\ and\ \citenamefont
  {{Vaillancourt}}}]{hildebrand99}%
  \BibitemOpen
  \bibfield  {author} {\bibinfo {author} {\bibfnamefont {R.~H.}\ \bibnamefont
  {{Hildebrand}}}, \bibinfo {author} {\bibfnamefont {J.~L.}\ \bibnamefont
  {{Dotson}}}, \bibinfo {author} {\bibfnamefont {C.~D.}\ \bibnamefont
  {{Dowell}}}, \bibinfo {author} {\bibfnamefont {D.~A.}\ \bibnamefont
  {{Schleuning}}}, \ and\ \bibinfo {author} {\bibfnamefont {J.~E.}\
  \bibnamefont {{Vaillancourt}}},\ }\href {\doibase 10.1086/307142} {\bibfield
  {journal} {\bibinfo  {journal} {\apj}\ }\textbf {\bibinfo {volume} {516}},\
  \bibinfo {pages} {834} (\bibinfo {year} {1999})}\BibitemShut {NoStop}%
\bibitem [{\citenamefont {{Draine}}(2004)}]{draine04}%
  \BibitemOpen
  \bibfield  {author} {\bibinfo {author} {\bibfnamefont {B.~T.}\ \bibnamefont
  {{Draine}}},\ }in\ \href@noop {} {\emph {\bibinfo {booktitle} {The Cold
  Universe, Saas-Fee Advanced Course 32, Springer-Verlag, 308 pages, 129
  figures, Lecture Notes 2002 of the Swiss Society for Astronomy and
  Astrophysics (SSAA), Springer, 2004. Edited by A.W. Blain, F. Combes, B.T.
  Draine, D. Pfenniger and Y. Revaz, ISBN 354040838x, p. 213}}},\ \bibinfo
  {editor} {edited by\ \bibinfo {editor} {\bibfnamefont {A.~W.}\ \bibnamefont
  {{Blain}}}, \bibinfo {editor} {\bibfnamefont {F.}~\bibnamefont {{Combes}}},
  \bibinfo {editor} {\bibfnamefont {B.~T.}\ \bibnamefont {{Draine}}}, \bibinfo
  {editor} {\bibfnamefont {D.}~\bibnamefont {{Pfenniger}}}, \ and\ \bibinfo
  {editor} {\bibfnamefont {Y.}~\bibnamefont {{Revaz}}}}\ (\bibinfo {year}
  {2004})\ p.\ \bibinfo {pages} {213},\ \Eprint
  {http://arxiv.org/abs/astro-ph/0304488} {astro-ph/0304488} \BibitemShut
  {NoStop}%
\bibitem [{\citenamefont {{Martin}}(2007)}]{martin07}%
  \BibitemOpen
  \bibfield  {author} {\bibinfo {author} {\bibfnamefont {P.~G.}\ \bibnamefont
  {{Martin}}},\ }in\ \href {\doibase 10.1051/eas:2007011} {\emph {\bibinfo
  {booktitle} {EAS Publications Series}}},\ \bibinfo {series} {EAS Publications
  Series}, Vol.~\bibinfo {volume} {23},\ \bibinfo {editor} {edited by\ \bibinfo
  {editor} {\bibfnamefont {M.-A.}\ \bibnamefont {{Miville-Desch{\^e}nes}}}\
  and\ \bibinfo {editor} {\bibfnamefont {F.}~\bibnamefont {{Boulanger}}}}\
  (\bibinfo {year} {2007})\ pp.\ \bibinfo {pages} {165--188},\ \Eprint
  {http://arxiv.org/abs/astro-ph/0606430} {astro-ph/0606430} \BibitemShut
  {NoStop}%
\bibitem [{\citenamefont {{Beno{\^i}t}}\ \emph {et~al.}(2004)\citenamefont
  {{Beno{\^i}t}}, \citenamefont {{Ade}}, \citenamefont {{Amblard}},
  \citenamefont {{Ansari}}, \citenamefont {{Aubourg}}, \citenamefont
  {{Bargot}}, \citenamefont {{Bartlett}}, \citenamefont {{Bernard}},
  \citenamefont {{Bhatia}}, \citenamefont {{Blanchard}}, \citenamefont
  {{Bock}}, \citenamefont {{Boscaleri}}, \citenamefont {{Bouchet}},
  \citenamefont {{Bourrachot}}, \citenamefont {{Camus}}, \citenamefont
  {{Couchot}}, \citenamefont {{de Bernardis}}, \citenamefont {{Delabrouille}},
  \citenamefont {{D{\'e}sert}}, \citenamefont {{Dor{\'e}}}, \citenamefont
  {{Douspis}}, \citenamefont {{Dumoulin}}, \citenamefont {{Dupac}},
  \citenamefont {{Filliatre}}, \citenamefont {{Fosalba}}, \citenamefont
  {{Ganga}}, \citenamefont {{Gannaway}}, \citenamefont {{Gautier}},
  \citenamefont {{Giard}}, \citenamefont {{Giraud-H{\'e}raud}}, \citenamefont
  {{Gispert}}, \citenamefont {{Guglielmi}}, \citenamefont {{Hamilton}},
  \citenamefont {{Hanany}}, \citenamefont {{Henrot-Versill{\'e}}},
  \citenamefont {{Kaplan}}, \citenamefont {{Lagache}}, \citenamefont
  {{Lamarre}}, \citenamefont {{Lange}}, \citenamefont
  {{Mac{\'{\i}}as-P{\'e}rez}}, \citenamefont {{Madet}}, \citenamefont
  {{Maffei}}, \citenamefont {{Magneville}}, \citenamefont {{Marrone}},
  \citenamefont {{Masi}}, \citenamefont {{Mayet}}, \citenamefont {{Murphy}},
  \citenamefont {{Naraghi}}, \citenamefont {{Nati}}, \citenamefont
  {{Patanchon}}, \citenamefont {{Perrin}}, \citenamefont {{Piat}},
  \citenamefont {{Ponthieu}}, \citenamefont {{Prunet}}, \citenamefont
  {{Puget}}, \citenamefont {{Renault}}, \citenamefont {{Rosset}}, \citenamefont
  {{Santos}}, \citenamefont {{Starobinsky}}, \citenamefont {{Strukov}},
  \citenamefont {{Sudiwala}}, \citenamefont {{Teyssier}}, \citenamefont
  {{Tristram}}, \citenamefont {{Tucker}}, \citenamefont {{Vanel}},
  \citenamefont {{Vibert}}, \citenamefont {{Wakui}},\ and\ \citenamefont
  {{Yvon}}}]{archeopsdust}%
  \BibitemOpen
  \bibfield  {author} {\bibinfo {author} {\bibfnamefont {A.}~\bibnamefont
  {{Beno{\^i}t}}}, \bibinfo {author} {\bibfnamefont {P.}~\bibnamefont {{Ade}}},
  \bibinfo {author} {\bibfnamefont {A.}~\bibnamefont {{Amblard}}}, \bibinfo
  {author} {\bibfnamefont {R.}~\bibnamefont {{Ansari}}}, \bibinfo {author}
  {\bibfnamefont {{\'E}.}~\bibnamefont {{Aubourg}}}, \bibinfo {author}
  {\bibfnamefont {S.}~\bibnamefont {{Bargot}}}, \bibinfo {author}
  {\bibfnamefont {J.~G.}\ \bibnamefont {{Bartlett}}}, \bibinfo {author}
  {\bibfnamefont {J.-P.}\ \bibnamefont {{Bernard}}}, \bibinfo {author}
  {\bibfnamefont {R.~S.}\ \bibnamefont {{Bhatia}}}, \bibinfo {author}
  {\bibfnamefont {A.}~\bibnamefont {{Blanchard}}}, \bibinfo {author}
  {\bibfnamefont {J.~J.}\ \bibnamefont {{Bock}}}, \bibinfo {author}
  {\bibfnamefont {A.}~\bibnamefont {{Boscaleri}}}, \bibinfo {author}
  {\bibfnamefont {F.~R.}\ \bibnamefont {{Bouchet}}}, \bibinfo {author}
  {\bibfnamefont {A.}~\bibnamefont {{Bourrachot}}}, \bibinfo {author}
  {\bibfnamefont {P.}~\bibnamefont {{Camus}}}, \bibinfo {author} {\bibfnamefont
  {F.}~\bibnamefont {{Couchot}}}, \bibinfo {author} {\bibfnamefont
  {P.}~\bibnamefont {{de Bernardis}}}, \bibinfo {author} {\bibfnamefont
  {J.}~\bibnamefont {{Delabrouille}}}, \bibinfo {author} {\bibfnamefont
  {F.-X.}\ \bibnamefont {{D{\'e}sert}}}, \bibinfo {author} {\bibfnamefont
  {O.}~\bibnamefont {{Dor{\'e}}}}, \bibinfo {author} {\bibfnamefont
  {M.}~\bibnamefont {{Douspis}}}, \bibinfo {author} {\bibfnamefont
  {L.}~\bibnamefont {{Dumoulin}}}, \bibinfo {author} {\bibfnamefont
  {X.}~\bibnamefont {{Dupac}}}, \bibinfo {author} {\bibfnamefont
  {P.}~\bibnamefont {{Filliatre}}}, \bibinfo {author} {\bibfnamefont
  {P.}~\bibnamefont {{Fosalba}}}, \bibinfo {author} {\bibfnamefont
  {K.}~\bibnamefont {{Ganga}}}, \bibinfo {author} {\bibfnamefont
  {F.}~\bibnamefont {{Gannaway}}}, \bibinfo {author} {\bibfnamefont
  {B.}~\bibnamefont {{Gautier}}}, \bibinfo {author} {\bibfnamefont
  {M.}~\bibnamefont {{Giard}}}, \bibinfo {author} {\bibfnamefont
  {Y.}~\bibnamefont {{Giraud-H{\'e}raud}}}, \bibinfo {author} {\bibfnamefont
  {R.}~\bibnamefont {{Gispert}}}, \bibinfo {author} {\bibfnamefont
  {L.}~\bibnamefont {{Guglielmi}}}, \bibinfo {author} {\bibfnamefont {J.-C.}\
  \bibnamefont {{Hamilton}}}, \bibinfo {author} {\bibfnamefont
  {S.}~\bibnamefont {{Hanany}}}, \bibinfo {author} {\bibfnamefont
  {S.}~\bibnamefont {{Henrot-Versill{\'e}}}}, \bibinfo {author} {\bibfnamefont
  {J.}~\bibnamefont {{Kaplan}}}, \bibinfo {author} {\bibfnamefont
  {G.}~\bibnamefont {{Lagache}}}, \bibinfo {author} {\bibfnamefont {J.-M.}\
  \bibnamefont {{Lamarre}}}, \bibinfo {author} {\bibfnamefont {A.~E.}\
  \bibnamefont {{Lange}}}, \bibinfo {author} {\bibfnamefont {J.~F.}\
  \bibnamefont {{Mac{\'{\i}}as-P{\'e}rez}}}, \bibinfo {author} {\bibfnamefont
  {K.}~\bibnamefont {{Madet}}}, \bibinfo {author} {\bibfnamefont
  {B.}~\bibnamefont {{Maffei}}}, \bibinfo {author} {\bibfnamefont
  {C.}~\bibnamefont {{Magneville}}}, \bibinfo {author} {\bibfnamefont {D.~P.}\
  \bibnamefont {{Marrone}}}, \bibinfo {author} {\bibfnamefont {S.}~\bibnamefont
  {{Masi}}}, \bibinfo {author} {\bibfnamefont {F.}~\bibnamefont {{Mayet}}},
  \bibinfo {author} {\bibfnamefont {A.}~\bibnamefont {{Murphy}}}, \bibinfo
  {author} {\bibfnamefont {F.}~\bibnamefont {{Naraghi}}}, \bibinfo {author}
  {\bibfnamefont {F.}~\bibnamefont {{Nati}}}, \bibinfo {author} {\bibfnamefont
  {G.}~\bibnamefont {{Patanchon}}}, \bibinfo {author} {\bibfnamefont
  {G.}~\bibnamefont {{Perrin}}}, \bibinfo {author} {\bibfnamefont
  {M.}~\bibnamefont {{Piat}}}, \bibinfo {author} {\bibfnamefont
  {N.}~\bibnamefont {{Ponthieu}}}, \bibinfo {author} {\bibfnamefont
  {S.}~\bibnamefont {{Prunet}}}, \bibinfo {author} {\bibfnamefont {J.-L.}\
  \bibnamefont {{Puget}}}, \bibinfo {author} {\bibfnamefont {C.}~\bibnamefont
  {{Renault}}}, \bibinfo {author} {\bibfnamefont {C.}~\bibnamefont {{Rosset}}},
  \bibinfo {author} {\bibfnamefont {D.}~\bibnamefont {{Santos}}}, \bibinfo
  {author} {\bibfnamefont {A.}~\bibnamefont {{Starobinsky}}}, \bibinfo {author}
  {\bibfnamefont {I.}~\bibnamefont {{Strukov}}}, \bibinfo {author}
  {\bibfnamefont {R.~V.}\ \bibnamefont {{Sudiwala}}}, \bibinfo {author}
  {\bibfnamefont {R.}~\bibnamefont {{Teyssier}}}, \bibinfo {author}
  {\bibfnamefont {M.}~\bibnamefont {{Tristram}}}, \bibinfo {author}
  {\bibfnamefont {C.}~\bibnamefont {{Tucker}}}, \bibinfo {author}
  {\bibfnamefont {J.-C.}\ \bibnamefont {{Vanel}}}, \bibinfo {author}
  {\bibfnamefont {D.}~\bibnamefont {{Vibert}}}, \bibinfo {author}
  {\bibfnamefont {E.}~\bibnamefont {{Wakui}}}, \ and\ \bibinfo {author}
  {\bibfnamefont {D.}~\bibnamefont {{Yvon}}},\ }\href {\doibase
  10.1051/0004-6361:20040042} {\bibfield  {journal} {\bibinfo  {journal}
  {\aap}\ }\textbf {\bibinfo {volume} {424}},\ \bibinfo {pages} {571} (\bibinfo
  {year} {2004})},\ \Eprint {http://arxiv.org/abs/astro-ph/0306222}
  {astro-ph/0306222} \BibitemShut {NoStop}%
\bibitem [{\citenamefont {{Page}}\ \emph {et~al.}(2007)\citenamefont {{Page}},
  \citenamefont {{Hinshaw}}, \citenamefont {{Komatsu}}, \citenamefont
  {{Nolta}}, \citenamefont {{Spergel}}, \citenamefont {{Bennett}},
  \citenamefont {{Barnes}}, \citenamefont {{Bean}}, \citenamefont {{Dor{\'e}}},
  \citenamefont {{Dunkley}}, \citenamefont {{Halpern}}, \citenamefont {{Hill}},
  \citenamefont {{Jarosik}}, \citenamefont {{Kogut}}, \citenamefont {{Limon}},
  \citenamefont {{Meyer}}, \citenamefont {{Odegard}}, \citenamefont {{Peiris}},
  \citenamefont {{Tucker}}, \citenamefont {{Verde}} \emph {et~al.}}]{page07}%
  \BibitemOpen
  \bibfield  {author} {\bibinfo {author} {\bibfnamefont {L.}~\bibnamefont
  {{Page}}}, \bibinfo {author} {\bibfnamefont {G.}~\bibnamefont {{Hinshaw}}},
  \bibinfo {author} {\bibfnamefont {E.}~\bibnamefont {{Komatsu}}}, \bibinfo
  {author} {\bibfnamefont {M.~R.}\ \bibnamefont {{Nolta}}}, \bibinfo {author}
  {\bibfnamefont {D.~N.}\ \bibnamefont {{Spergel}}}, \bibinfo {author}
  {\bibfnamefont {C.~L.}\ \bibnamefont {{Bennett}}}, \bibinfo {author}
  {\bibfnamefont {C.}~\bibnamefont {{Barnes}}}, \bibinfo {author}
  {\bibfnamefont {R.}~\bibnamefont {{Bean}}}, \bibinfo {author} {\bibfnamefont
  {O.}~\bibnamefont {{Dor{\'e}}}}, \bibinfo {author} {\bibfnamefont
  {J.}~\bibnamefont {{Dunkley}}}, \bibinfo {author} {\bibfnamefont
  {M.}~\bibnamefont {{Halpern}}}, \bibinfo {author} {\bibfnamefont {R.~S.}\
  \bibnamefont {{Hill}}}, \bibinfo {author} {\bibfnamefont {N.}~\bibnamefont
  {{Jarosik}}}, \bibinfo {author} {\bibfnamefont {A.}~\bibnamefont {{Kogut}}},
  \bibinfo {author} {\bibfnamefont {M.}~\bibnamefont {{Limon}}}, \bibinfo
  {author} {\bibfnamefont {S.~S.}\ \bibnamefont {{Meyer}}}, \bibinfo {author}
  {\bibfnamefont {N.}~\bibnamefont {{Odegard}}}, \bibinfo {author}
  {\bibfnamefont {H.~V.}\ \bibnamefont {{Peiris}}}, \bibinfo {author}
  {\bibfnamefont {G.~S.}\ \bibnamefont {{Tucker}}}, \bibinfo {author}
  {\bibfnamefont {L.}~\bibnamefont {{Verde}}},  \emph {et~al.},\ }\href
  {\doibase 10.1086/513699} {\bibfield  {journal} {\bibinfo  {journal} {\apjs}\
  }\textbf {\bibinfo {volume} {170}},\ \bibinfo {pages} {335} (\bibinfo {year}
  {2007})},\ \Eprint {http://arxiv.org/abs/astro-ph/0603450} {astro-ph/0603450}
  \BibitemShut {NoStop}%
\bibitem [{\citenamefont {{\textsc{Bicep2} Collaboration
  II}}(2014)}]{b2instpap14}%
  \BibitemOpen
  \bibfield  {author} {\bibinfo {author} {\bibnamefont {{\textsc{Bicep2}
  Collaboration II}}},\ }\href {\doibase 10.1088/0004-637X/792/1/62} {\bibfield
   {journal} {\bibinfo  {journal} {\apj}\ }\textbf {\bibinfo {volume} {792}},\
  \bibinfo {eid} {62} (\bibinfo {year} {2014})}\BibitemShut {NoStop}%
\bibitem [{\citenamefont {{\textsc{Bicep2} Collaboration
  I}}(2014)}]{biceptwoI}%
  \BibitemOpen
  \bibfield  {author} {\bibinfo {author} {\bibnamefont {{\textsc{Bicep2}
  Collaboration I}}},\ }\href {\doibase 10.1103/PhysRevLett.112.241101}
  {\bibfield  {journal} {\bibinfo  {journal} {Phys. Rev. Lett.}\ }\textbf
  {\bibinfo {volume} {112}},\ \bibinfo {pages} {241101} (\bibinfo {year}
  {2014})}\BibitemShut {NoStop}%
\bibitem [{\citenamefont {{Dunkley}}\ \emph {et~al.}(2009)\citenamefont
  {{Dunkley}}, \citenamefont {{Amblard}}, \citenamefont {{Baccigalupi}},
  \citenamefont {{Betoule}}, \citenamefont {{Chuss}}, \citenamefont {{Cooray}},
  \citenamefont {{Delabrouille}}, \citenamefont {{Dickinson}}, \citenamefont
  {{Dobler}}, \citenamefont {{Dotson}}, \citenamefont {{Eriksen}},
  \citenamefont {{Finkbeiner}}, \citenamefont {{Fixsen}}, \citenamefont
  {{Fosalba}}, \citenamefont {{Fraisse}}, \citenamefont {{Hirata}},
  \citenamefont {{Kogut}}, \citenamefont {{Kristiansen}}, \citenamefont
  {{Lawrence}}, \citenamefont {{Magalh\~{a}es}}, \citenamefont
  {{Miville-Deschenes}} \emph {et~al.}}]{dunkley08}%
  \BibitemOpen
  \bibfield  {author} {\bibinfo {author} {\bibfnamefont {J.}~\bibnamefont
  {{Dunkley}}}, \bibinfo {author} {\bibfnamefont {A.}~\bibnamefont
  {{Amblard}}}, \bibinfo {author} {\bibfnamefont {C.}~\bibnamefont
  {{Baccigalupi}}}, \bibinfo {author} {\bibfnamefont {M.}~\bibnamefont
  {{Betoule}}}, \bibinfo {author} {\bibfnamefont {D.}~\bibnamefont {{Chuss}}},
  \bibinfo {author} {\bibfnamefont {A.}~\bibnamefont {{Cooray}}}, \bibinfo
  {author} {\bibfnamefont {J.}~\bibnamefont {{Delabrouille}}}, \bibinfo
  {author} {\bibfnamefont {C.}~\bibnamefont {{Dickinson}}}, \bibinfo {author}
  {\bibfnamefont {G.}~\bibnamefont {{Dobler}}}, \bibinfo {author}
  {\bibfnamefont {J.}~\bibnamefont {{Dotson}}}, \bibinfo {author}
  {\bibfnamefont {H.~K.}\ \bibnamefont {{Eriksen}}}, \bibinfo {author}
  {\bibfnamefont {D.}~\bibnamefont {{Finkbeiner}}}, \bibinfo {author}
  {\bibfnamefont {D.}~\bibnamefont {{Fixsen}}}, \bibinfo {author}
  {\bibfnamefont {P.}~\bibnamefont {{Fosalba}}}, \bibinfo {author}
  {\bibfnamefont {A.}~\bibnamefont {{Fraisse}}}, \bibinfo {author}
  {\bibfnamefont {C.}~\bibnamefont {{Hirata}}}, \bibinfo {author}
  {\bibfnamefont {A.}~\bibnamefont {{Kogut}}}, \bibinfo {author} {\bibfnamefont
  {J.}~\bibnamefont {{Kristiansen}}}, \bibinfo {author} {\bibfnamefont
  {C.}~\bibnamefont {{Lawrence}}}, \bibinfo {author} {\bibfnamefont {A.~M.}\
  \bibnamefont {{Magalh\~{a}es}}}, \bibinfo {author} {\bibfnamefont {M.~A.}\
  \bibnamefont {{Miville-Deschenes}}},  \emph {et~al.},\ }\href {\doibase
  10.1063/1.3160888} {\bibfield  {journal} {\bibinfo  {journal} {AIP Conf.\
  Proc.}\ }\textbf {\bibinfo {volume} {1141}},\ \bibinfo {pages} {222}
  (\bibinfo {year} {2009})},\ \Eprint {http://arxiv.org/abs/0811.3915}
  {arXiv:0811.3915} \BibitemShut {NoStop}%
\bibitem [{\citenamefont {{Delabrouille}}\ \emph {et~al.}(2013)\citenamefont
  {{Delabrouille}}, \citenamefont {{Betoule}}, \citenamefont {{Melin}},
  \citenamefont {{Miville-Desch{\^e}nes}}, \citenamefont {{Gonzalez-Nuevo}},
  \citenamefont {{Le Jeune}}, \citenamefont {{Castex}}, \citenamefont {{de
  Zotti}}, \citenamefont {{Basak}}, \citenamefont {{Ashdown}}, \citenamefont
  {{Aumont}}, \citenamefont {{Baccigalupi}}, \citenamefont {{Banday}},
  \citenamefont {{Bernard}}, \citenamefont {{Bouchet}}, \citenamefont
  {{Clements}}, \citenamefont {{da Silva}}, \citenamefont {{Dickinson}},
  \citenamefont {{Dodu}}, \citenamefont {{Dolag}} \emph
  {et~al.}}]{delabrouille13}%
  \BibitemOpen
  \bibfield  {author} {\bibinfo {author} {\bibfnamefont {J.}~\bibnamefont
  {{Delabrouille}}}, \bibinfo {author} {\bibfnamefont {M.}~\bibnamefont
  {{Betoule}}}, \bibinfo {author} {\bibfnamefont {J.-B.}\ \bibnamefont
  {{Melin}}}, \bibinfo {author} {\bibfnamefont {M.-A.}\ \bibnamefont
  {{Miville-Desch{\^e}nes}}}, \bibinfo {author} {\bibfnamefont
  {J.}~\bibnamefont {{Gonzalez-Nuevo}}}, \bibinfo {author} {\bibfnamefont
  {M.}~\bibnamefont {{Le Jeune}}}, \bibinfo {author} {\bibfnamefont
  {G.}~\bibnamefont {{Castex}}}, \bibinfo {author} {\bibfnamefont
  {G.}~\bibnamefont {{de Zotti}}}, \bibinfo {author} {\bibfnamefont
  {S.}~\bibnamefont {{Basak}}}, \bibinfo {author} {\bibfnamefont
  {M.}~\bibnamefont {{Ashdown}}}, \bibinfo {author} {\bibfnamefont
  {J.}~\bibnamefont {{Aumont}}}, \bibinfo {author} {\bibfnamefont
  {C.}~\bibnamefont {{Baccigalupi}}}, \bibinfo {author} {\bibfnamefont {A.~J.}\
  \bibnamefont {{Banday}}}, \bibinfo {author} {\bibfnamefont {J.-P.}\
  \bibnamefont {{Bernard}}}, \bibinfo {author} {\bibfnamefont {F.~R.}\
  \bibnamefont {{Bouchet}}}, \bibinfo {author} {\bibfnamefont {D.~L.}\
  \bibnamefont {{Clements}}}, \bibinfo {author} {\bibfnamefont
  {A.}~\bibnamefont {{da Silva}}}, \bibinfo {author} {\bibfnamefont
  {C.}~\bibnamefont {{Dickinson}}}, \bibinfo {author} {\bibfnamefont
  {F.}~\bibnamefont {{Dodu}}}, \bibinfo {author} {\bibfnamefont
  {K.}~\bibnamefont {{Dolag}}},  \emph {et~al.},\ }\href {\doibase
  10.1051/0004-6361/201220019} {\bibfield  {journal} {\bibinfo  {journal}
  {\aap}\ }\textbf {\bibinfo {volume} {553}},\ \bibinfo {eid} {A96} (\bibinfo
  {year} {2013})},\ \Eprint {http://arxiv.org/abs/1207.3675} {arXiv:1207.3675}
  \BibitemShut {NoStop}%
\bibitem [{\citenamefont {{Planck Collaboration Int.\
  XIX}}(2014)}]{planckiXIX}%
  \BibitemOpen
  \bibfield  {author} {\bibinfo {author} {\bibnamefont {{Planck Collaboration
  Int.\ XIX}}},\ }\href@noop {} {\  (\bibinfo {year} {2014})},\ \Eprint
  {http://arxiv.org/abs/1405.0871v1} {arXiv:1405.0871v1} \BibitemShut {NoStop}%
\bibitem [{\citenamefont {{Planck Collaboration Int.
  XXII}}(2014)}]{planckiXXII}%
  \BibitemOpen
  \bibfield  {author} {\bibinfo {author} {\bibnamefont {{Planck Collaboration
  Int. XXII}}},\ }\href@noop {} {\bibfield  {journal} {\bibinfo  {journal}
  {\aap}\ } (\bibinfo {year} {2014})},\ \Eprint
  {http://arxiv.org/abs/1405.0874} {arXiv:1405.0874} \BibitemShut {NoStop}%
\bibitem [{\citenamefont {{Flauger}}\ \emph {et~al.}(2014)\citenamefont
  {{Flauger}}, \citenamefont {{Hill}},\ and\ \citenamefont
  {{Spergel}}}]{flauger14}%
  \BibitemOpen
  \bibfield  {author} {\bibinfo {author} {\bibfnamefont {R.}~\bibnamefont
  {{Flauger}}}, \bibinfo {author} {\bibfnamefont {J.~C.}\ \bibnamefont
  {{Hill}}}, \ and\ \bibinfo {author} {\bibfnamefont {D.~N.}\ \bibnamefont
  {{Spergel}}},\ }\href {\doibase 10.1088/1475-7516/2014/08/039} {\bibfield
  {journal} {\bibinfo  {journal} {\jcap}\ }\textbf {\bibinfo {volume} {8}},\
  \bibinfo {eid} {039} (\bibinfo {year} {2014})},\ \Eprint
  {http://arxiv.org/abs/1405.7351} {arXiv:1405.7351} \BibitemShut {NoStop}%
\bibitem [{\citenamefont {{Mortonson}}\ and\ \citenamefont
  {{Seljak}}(2014)}]{mortonson14}%
  \BibitemOpen
  \bibfield  {author} {\bibinfo {author} {\bibfnamefont {M.~J.}\ \bibnamefont
  {{Mortonson}}}\ and\ \bibinfo {author} {\bibfnamefont {U.}~\bibnamefont
  {{Seljak}}},\ }\href {\doibase 10.1088/1475-7516/2014/10/035} {\bibfield
  {journal} {\bibinfo  {journal} {\jcap}\ }\textbf {\bibinfo {volume} {10}},\
  \bibinfo {eid} {035} (\bibinfo {year} {2014})},\ \Eprint
  {http://arxiv.org/abs/1405.5857} {arXiv:1405.5857} \BibitemShut {NoStop}%
\bibitem [{\citenamefont {{Planck Collaboration Int.\
  XXX}}(2014)}]{planckiXXX}%
  \BibitemOpen
  \bibfield  {author} {\bibinfo {author} {\bibnamefont {{Planck Collaboration
  Int.\ XXX}}},\ }\href@noop {} {\bibfield  {journal} {\bibinfo  {journal}
  {\aap}\ } (\bibinfo {year} {2014})},\ \Eprint
  {http://arxiv.org/abs/1409.5738} {arXiv:1409.5738} \BibitemShut {NoStop}%
\bibitem [{\citenamefont {{\keckarray\ and \textsc{Bicep2} Collaborations
  V}}()}]{biceptwoV}%
  \BibitemOpen
  \bibfield  {author} {\bibinfo {author} {\bibnamefont {{\keckarray\ and
  \textsc{Bicep2} Collaborations V}}},\ }\href@noop {} {\bibinfo  {journal}
  {submitted to \apj}\ }\BibitemShut {NoStop}%
\bibitem [{Note3()}]{Note3}%
  \BibitemOpen
\bibfield  {journal} {  }\bibinfo {note} {\protect \url
  {http://archives.esac.esa.int/pla2}}\BibitemShut {NoStop}%
\bibitem [{\citenamefont {{Planck Collaboration C01}}(2015)}]{planck2014-a01}%
  \BibitemOpen
  \bibfield  {author} {\bibinfo {author} {\bibnamefont {{Planck Collaboration
  C01}}},\ }\href@noop {} {\bibfield  {journal} {\bibinfo  {journal} {in
  preparation}\ } (\bibinfo {year} {2015})}\BibitemShut {NoStop}%
\bibitem [{\citenamefont {{Planck Collaboration C14}}(2015)}]{planck2014-a14}%
  \BibitemOpen
  \bibfield  {author} {\bibinfo {author} {\bibnamefont {{Planck Collaboration
  C14}}},\ }\href@noop {} {\bibfield  {journal} {\bibinfo  {journal} {in
  preparation}\ } (\bibinfo {year} {2015})}\BibitemShut {NoStop}%
\bibitem [{\citenamefont {{Planck Collaboration IV}}(2014)}]{planck2013IV}%
  \BibitemOpen
  \bibfield  {author} {\bibinfo {author} {\bibnamefont {{Planck Collaboration
  IV}}},\ }\href@noop {} {\bibfield  {journal} {\bibinfo  {journal} {\aap}\
  }\textbf {\bibinfo {volume} {571}},\ \bibinfo {pages} {A4} (\bibinfo {year}
  {2014})}\BibitemShut {NoStop}%
\bibitem [{\citenamefont {{Planck Collaboration VII}}(2014)}]{planck2013VII}%
  \BibitemOpen
  \bibfield  {author} {\bibinfo {author} {\bibnamefont {{Planck Collaboration
  VII}}},\ }\href@noop {} {\bibfield  {journal} {\bibinfo  {journal} {\aap}\
  }\textbf {\bibinfo {volume} {571}},\ \bibinfo {pages} {A7} (\bibinfo {year}
  {2014})}\BibitemShut {NoStop}%
\bibitem [{Note4()}]{Note4}%
  \BibitemOpen
  \bibinfo {note} {\protect \url {http://healpix.sourceforge.net/}}\BibitemShut
  {NoStop}%
\bibitem [{\citenamefont {{G{\'o}rski}}\ \emph {et~al.}(2005)\citenamefont
  {{G{\'o}rski}}, \citenamefont {{Hivon}}, \citenamefont {{Banday}},
  \citenamefont {{Wandelt}}, \citenamefont {{Hansen}}, \citenamefont
  {{Reinecke}},\ and\ \citenamefont {{Bartelmann}}}]{healpix}%
  \BibitemOpen
  \bibfield  {author} {\bibinfo {author} {\bibfnamefont {K.~M.}\ \bibnamefont
  {{G{\'o}rski}}}, \bibinfo {author} {\bibfnamefont {E.}~\bibnamefont
  {{Hivon}}}, \bibinfo {author} {\bibfnamefont {A.~J.}\ \bibnamefont
  {{Banday}}}, \bibinfo {author} {\bibfnamefont {B.~D.}\ \bibnamefont
  {{Wandelt}}}, \bibinfo {author} {\bibfnamefont {F.~K.}\ \bibnamefont
  {{Hansen}}}, \bibinfo {author} {\bibfnamefont {M.}~\bibnamefont
  {{Reinecke}}}, \ and\ \bibinfo {author} {\bibfnamefont {M.}~\bibnamefont
  {{Bartelmann}}},\ }\href {\doibase 10.1086/427976} {\bibfield  {journal}
  {\bibinfo  {journal} {\apj}\ }\textbf {\bibinfo {volume} {622}},\ \bibinfo
  {pages} {759} (\bibinfo {year} {2005})},\ \Eprint
  {http://arxiv.org/abs/astro-ph/0409513} {astro-ph/0409513} \BibitemShut
  {NoStop}%
\bibitem [{Note5()}]{Note5}%
  \BibitemOpen
  \bibinfo {note} {With parameters taken from {\protect \it Planck}~\cite
  {planck2013XVI}.}\BibitemShut {Stop}%
\bibitem [{\citenamefont {{Tristram}}\ \emph {et~al.}(2005)\citenamefont
  {{Tristram}}, \citenamefont {{Mac{\'{\i}}as-P{\'e}rez}}, \citenamefont
  {{Renault}},\ and\ \citenamefont {{Santos}}}]{tristram2005}%
  \BibitemOpen
  \bibfield  {author} {\bibinfo {author} {\bibfnamefont {M.}~\bibnamefont
  {{Tristram}}}, \bibinfo {author} {\bibfnamefont {J.~F.}\ \bibnamefont
  {{Mac{\'{\i}}as-P{\'e}rez}}}, \bibinfo {author} {\bibfnamefont
  {C.}~\bibnamefont {{Renault}}}, \ and\ \bibinfo {author} {\bibfnamefont
  {D.}~\bibnamefont {{Santos}}},\ }\href {\doibase
  10.1111/j.1365-2966.2005.08760.x} {\bibfield  {journal} {\bibinfo  {journal}
  {\mnras}\ }\textbf {\bibinfo {volume} {358}},\ \bibinfo {pages} {833}
  (\bibinfo {year} {2005})},\ \Eprint
  {http://arxiv.org/abs/arXiv:astro-ph/0405575} {arXiv:astro-ph/0405575}
  \BibitemShut {NoStop}%
\bibitem [{\citenamefont {Preece}(2011)}]{Preece11}%
  \BibitemOpen
  \bibfield  {author} {\bibinfo {author} {\bibfnamefont {M.}~\bibnamefont
  {Preece}},\ }\emph {\bibinfo {title} {{Analysis of Cosmic Microwave
  Background Polarisation}}},\ \href@noop {} {Ph.D. thesis},\ \bibinfo
  {school} {The University of Manchester, UK} (\bibinfo {year}
  {2011})\BibitemShut {NoStop}%
\bibitem [{\citenamefont {{Hamimeche}}\ and\ \citenamefont
  {{Lewis}}(2008)}]{hamimeche08}%
  \BibitemOpen
  \bibfield  {author} {\bibinfo {author} {\bibfnamefont {S.}~\bibnamefont
  {{Hamimeche}}}\ and\ \bibinfo {author} {\bibfnamefont {A.}~\bibnamefont
  {{Lewis}}},\ }\href {\doibase 10.1103/PhysRevD.77.103013} {\bibfield
  {journal} {\bibinfo  {journal} {\prd}\ }\textbf {\bibinfo {volume} {77}},\
  \bibinfo {eid} {103013} (\bibinfo {year} {2008})},\ \Eprint
  {http://arxiv.org/abs/0801.0554} {arXiv:0801.0554} \BibitemShut {NoStop}%
\bibitem [{\citenamefont {Barkats}\ \emph {et~al.}(2014)\citenamefont
  {Barkats}, \citenamefont {Aikin}, \citenamefont {Bischoff}, \citenamefont
  {Buder}, \citenamefont {Kaufman}, \citenamefont {Keating}, \citenamefont
  {Kovac}, \citenamefont {Su}, \citenamefont {Ade}, \citenamefont {Battle},
  \citenamefont {Bierman}, \citenamefont {Bock}, \citenamefont {Chiang},
  \citenamefont {Dowell}, \citenamefont {Duband}, \citenamefont {Filippini},
  \citenamefont {Hivon}, \citenamefont {Holzapfel}, \citenamefont {Hristov},
  \citenamefont {Jones} \emph {et~al.}}]{barkats14}%
  \BibitemOpen
  \bibfield  {author} {\bibinfo {author} {\bibfnamefont {D.}~\bibnamefont
  {Barkats}}, \bibinfo {author} {\bibfnamefont {R.}~\bibnamefont {Aikin}},
  \bibinfo {author} {\bibfnamefont {C.}~\bibnamefont {Bischoff}}, \bibinfo
  {author} {\bibfnamefont {I.}~\bibnamefont {Buder}}, \bibinfo {author}
  {\bibfnamefont {J.~P.}\ \bibnamefont {Kaufman}}, \bibinfo {author}
  {\bibfnamefont {B.~G.}\ \bibnamefont {Keating}}, \bibinfo {author}
  {\bibfnamefont {J.~M.}\ \bibnamefont {Kovac}}, \bibinfo {author}
  {\bibfnamefont {M.}~\bibnamefont {Su}}, \bibinfo {author} {\bibfnamefont
  {P.~A.~R.}\ \bibnamefont {Ade}}, \bibinfo {author} {\bibfnamefont {J.~O.}\
  \bibnamefont {Battle}}, \bibinfo {author} {\bibfnamefont {E.~M.}\
  \bibnamefont {Bierman}}, \bibinfo {author} {\bibfnamefont {J.~J.}\
  \bibnamefont {Bock}}, \bibinfo {author} {\bibfnamefont {H.~C.}\ \bibnamefont
  {Chiang}}, \bibinfo {author} {\bibfnamefont {C.~D.}\ \bibnamefont {Dowell}},
  \bibinfo {author} {\bibfnamefont {L.}~\bibnamefont {Duband}}, \bibinfo
  {author} {\bibfnamefont {J.}~\bibnamefont {Filippini}}, \bibinfo {author}
  {\bibfnamefont {E.~F.}\ \bibnamefont {Hivon}}, \bibinfo {author}
  {\bibfnamefont {W.~L.}\ \bibnamefont {Holzapfel}}, \bibinfo {author}
  {\bibfnamefont {V.~V.}\ \bibnamefont {Hristov}}, \bibinfo {author}
  {\bibfnamefont {W.~C.}\ \bibnamefont {Jones}},  \emph {et~al.},\ }\href
  {http://stacks.iop.org/0004-637X/783/i=2/a=67} {\bibfield  {journal}
  {\bibinfo  {journal} {\apj}\ }\textbf {\bibinfo {volume} {783}},\ \bibinfo
  {pages} {67} (\bibinfo {year} {2014})},\ \Eprint
  {http://arxiv.org/abs/1310.1422} {arXiv:1310.1422} \BibitemShut {NoStop}%
\bibitem [{\citenamefont {Lewis}\ and\ \citenamefont {Bridle}(2002)}]{cosmomc}%
  \BibitemOpen
  \bibfield  {author} {\bibinfo {author} {\bibfnamefont {A.}~\bibnamefont
  {Lewis}}\ and\ \bibinfo {author} {\bibfnamefont {S.}~\bibnamefont {Bridle}},\
  }\href@noop {} {\bibfield  {journal} {\bibinfo  {journal} {Phys. Rev.}\
  }\textbf {\bibinfo {volume} {D66}},\ \bibinfo {pages} {103511} (\bibinfo
  {year} {2002})},\ \Eprint {http://arxiv.org/abs/astro-ph/0205436}
  {astro-ph/0205436} \BibitemShut {NoStop}%
\bibitem [{\citenamefont {{Planck Collaboration Int.\
  XVII}}(2014)}]{planckiXVII}%
  \BibitemOpen
  \bibfield  {author} {\bibinfo {author} {\bibnamefont {{Planck Collaboration
  Int.\ XVII}}},\ }\href {\doibase 10.1051/0004-6361/201323270} {\bibfield
  {journal} {\bibinfo  {journal} {\aap}\ }\textbf {\bibinfo {volume} {566}},\
  \bibinfo {eid} {A55} (\bibinfo {year} {2014})},\ \Eprint
  {http://arxiv.org/abs/1312.5446} {arXiv:1312.5446} \BibitemShut {NoStop}%
\bibitem [{Note6()}]{Note6}%
  \BibitemOpen
  \bibinfo {note} {Note that this is the number evaluated at 353\protect \,GHz
  exactly---the equivalent number as integrated over the {\protect \it Planck}\
  353\protect \,GHz passband is 4.5\protect \,$\mu $K$^2$ and the mask used in
  PIP-XXX\ is somewhat different (larger) than the BICEP2/{\protect \it Keck}\
  mask used here.}\BibitemShut {Stop}%
\bibitem [{\citenamefont {{\textsc{Polarbear}
  Collaboration}}(2014)}]{polarbear14}%
  \BibitemOpen
  \bibfield  {author} {\bibinfo {author} {\bibnamefont {{\textsc{Polarbear}
  Collaboration}}},\ }\href {\doibase 10.1088/0004-637X/794/2/171} {\bibfield
  {journal} {\bibinfo  {journal} {\apj}\ }\textbf {\bibinfo {volume} {794}},\
  \bibinfo {eid} {171} (\bibinfo {year} {2014})},\ \Eprint
  {http://arxiv.org/abs/1403.2369} {arXiv:1403.2369} \BibitemShut {NoStop}%
\bibitem [{\citenamefont {{van Engelen}}\ \emph {et~al.}(2014)\citenamefont
  {{van Engelen}}, \citenamefont {{Sherwin}}, \citenamefont {{Sehgal}},
  \citenamefont {{Addison}}, \citenamefont {{Allison}}, \citenamefont
  {{Battaglia}}, \citenamefont {{de Bernardis}}, \citenamefont {{Calabrese}},
  \citenamefont {{Coughlin}}, \citenamefont {{Crichton}}, \citenamefont
  {{Bond}}, \citenamefont {{Datta}}, \citenamefont {{Dunner}}, \citenamefont
  {{Dunkley}}, \citenamefont {{Grace}}, \citenamefont {{Gralla}}, \citenamefont
  {{Hajian}}, \citenamefont {{Hasselfield}}, \citenamefont {{Henderson}},
  \citenamefont {{Hill}}, \citenamefont {{Hilton}}, \citenamefont {{Hincks}},
  \citenamefont {{Hlozek}}, \citenamefont {{Huffenberger}}, \citenamefont
  {{Hughes}}, \citenamefont {{Koopman}}, \citenamefont {{Kosowsky}},
  \citenamefont {{Louis}}, \citenamefont {{Lungu}}, \citenamefont
  {{Madhavacheril}}, \citenamefont {{Maurin}}, \citenamefont {{McMahon}},
  \citenamefont {{Moodley}}, \citenamefont {{Munson}}, \citenamefont {{Naess}},
  \citenamefont {{Nati}}, \citenamefont {{Newburgh}}, \citenamefont
  {{Niemack}}, \citenamefont {{Nolta}}, \citenamefont {{Page}}, \citenamefont
  {{Partridge}}, \citenamefont {{Pappas}}, \citenamefont {{Schmitt}},
  \citenamefont {{Sievers}}, \citenamefont {{Simon}}, \citenamefont
  {{Spergel}}, \citenamefont {{Staggs}}, \citenamefont {{Switzer}},
  \citenamefont {{Ward}},\ and\ \citenamefont {{Wollack}}}]{vanengelen14}%
  \BibitemOpen
  \bibfield  {author} {\bibinfo {author} {\bibfnamefont {A.}~\bibnamefont {{van
  Engelen}}}, \bibinfo {author} {\bibfnamefont {B.~D.}\ \bibnamefont
  {{Sherwin}}}, \bibinfo {author} {\bibfnamefont {N.}~\bibnamefont {{Sehgal}}},
  \bibinfo {author} {\bibfnamefont {G.~E.}\ \bibnamefont {{Addison}}}, \bibinfo
  {author} {\bibfnamefont {R.}~\bibnamefont {{Allison}}}, \bibinfo {author}
  {\bibfnamefont {N.}~\bibnamefont {{Battaglia}}}, \bibinfo {author}
  {\bibfnamefont {F.}~\bibnamefont {{de Bernardis}}}, \bibinfo {author}
  {\bibfnamefont {E.}~\bibnamefont {{Calabrese}}}, \bibinfo {author}
  {\bibfnamefont {K.}~\bibnamefont {{Coughlin}}}, \bibinfo {author}
  {\bibfnamefont {D.}~\bibnamefont {{Crichton}}}, \bibinfo {author}
  {\bibfnamefont {J.~R.}\ \bibnamefont {{Bond}}}, \bibinfo {author}
  {\bibfnamefont {R.}~\bibnamefont {{Datta}}}, \bibinfo {author} {\bibfnamefont
  {R.}~\bibnamefont {{Dunner}}}, \bibinfo {author} {\bibfnamefont
  {J.}~\bibnamefont {{Dunkley}}}, \bibinfo {author} {\bibfnamefont
  {E.}~\bibnamefont {{Grace}}}, \bibinfo {author} {\bibfnamefont
  {M.}~\bibnamefont {{Gralla}}}, \bibinfo {author} {\bibfnamefont
  {A.}~\bibnamefont {{Hajian}}}, \bibinfo {author} {\bibfnamefont
  {M.}~\bibnamefont {{Hasselfield}}}, \bibinfo {author} {\bibfnamefont
  {S.}~\bibnamefont {{Henderson}}}, \bibinfo {author} {\bibfnamefont {J.~C.}\
  \bibnamefont {{Hill}}}, \bibinfo {author} {\bibfnamefont {M.}~\bibnamefont
  {{Hilton}}}, \bibinfo {author} {\bibfnamefont {A.~D.}\ \bibnamefont
  {{Hincks}}}, \bibinfo {author} {\bibfnamefont {R.}~\bibnamefont {{Hlozek}}},
  \bibinfo {author} {\bibfnamefont {K.~M.}\ \bibnamefont {{Huffenberger}}},
  \bibinfo {author} {\bibfnamefont {J.~P.}\ \bibnamefont {{Hughes}}}, \bibinfo
  {author} {\bibfnamefont {B.}~\bibnamefont {{Koopman}}}, \bibinfo {author}
  {\bibfnamefont {A.}~\bibnamefont {{Kosowsky}}}, \bibinfo {author}
  {\bibfnamefont {T.}~\bibnamefont {{Louis}}}, \bibinfo {author} {\bibfnamefont
  {M.}~\bibnamefont {{Lungu}}}, \bibinfo {author} {\bibfnamefont
  {M.}~\bibnamefont {{Madhavacheril}}}, \bibinfo {author} {\bibfnamefont
  {L.}~\bibnamefont {{Maurin}}}, \bibinfo {author} {\bibfnamefont
  {J.}~\bibnamefont {{McMahon}}}, \bibinfo {author} {\bibfnamefont
  {K.}~\bibnamefont {{Moodley}}}, \bibinfo {author} {\bibfnamefont
  {C.}~\bibnamefont {{Munson}}}, \bibinfo {author} {\bibfnamefont
  {S.}~\bibnamefont {{Naess}}}, \bibinfo {author} {\bibfnamefont
  {F.}~\bibnamefont {{Nati}}}, \bibinfo {author} {\bibfnamefont
  {L.}~\bibnamefont {{Newburgh}}}, \bibinfo {author} {\bibfnamefont {M.~D.}\
  \bibnamefont {{Niemack}}}, \bibinfo {author} {\bibfnamefont {M.}~\bibnamefont
  {{Nolta}}}, \bibinfo {author} {\bibfnamefont {L.~A.}\ \bibnamefont {{Page}}},
  \bibinfo {author} {\bibfnamefont {B.}~\bibnamefont {{Partridge}}}, \bibinfo
  {author} {\bibfnamefont {C.}~\bibnamefont {{Pappas}}}, \bibinfo {author}
  {\bibfnamefont {B.~L.}\ \bibnamefont {{Schmitt}}}, \bibinfo {author}
  {\bibfnamefont {J.~L.}\ \bibnamefont {{Sievers}}}, \bibinfo {author}
  {\bibfnamefont {S.}~\bibnamefont {{Simon}}}, \bibinfo {author} {\bibfnamefont
  {D.~N.}\ \bibnamefont {{Spergel}}}, \bibinfo {author} {\bibfnamefont {S.~T.}\
  \bibnamefont {{Staggs}}}, \bibinfo {author} {\bibfnamefont {E.~R.}\
  \bibnamefont {{Switzer}}}, \bibinfo {author} {\bibfnamefont {J.~T.}\
  \bibnamefont {{Ward}}}, \ and\ \bibinfo {author} {\bibfnamefont {E.~J.}\
  \bibnamefont {{Wollack}}},\ }\href@noop {} {\bibfield  {journal} {\bibinfo
  {journal} {ArXiv e-prints}\ } (\bibinfo {year} {2014})},\ \Eprint
  {http://arxiv.org/abs/1412.0626} {arXiv:1412.0626} \BibitemShut {NoStop}%
\bibitem [{\citenamefont {{Hanson}}\ \emph {et~al.}(2013)\citenamefont
  {{Hanson}}, \citenamefont {{Hoover}}, \citenamefont {{Crites}}, \citenamefont
  {{Ade}}, \citenamefont {{Aird}}, \citenamefont {{Austermann}}, \citenamefont
  {{Beall}}, \citenamefont {{Bender}}, \citenamefont {{Benson}}, \citenamefont
  {{Bleem}}, \citenamefont {{Bock}}, \citenamefont {{Carlstrom}}, \citenamefont
  {{Chang}}, \citenamefont {{Chiang}}, \citenamefont {{Cho}}, \citenamefont
  {{Conley}}, \citenamefont {{Crawford}}, \citenamefont {{de Haan}},
  \citenamefont {{Dobbs}}, \citenamefont {{Everett}} \emph
  {et~al.}}]{hanson13}%
  \BibitemOpen
  \bibfield  {author} {\bibinfo {author} {\bibfnamefont {D.}~\bibnamefont
  {{Hanson}}}, \bibinfo {author} {\bibfnamefont {S.}~\bibnamefont {{Hoover}}},
  \bibinfo {author} {\bibfnamefont {A.}~\bibnamefont {{Crites}}}, \bibinfo
  {author} {\bibfnamefont {P.~A.~R.}\ \bibnamefont {{Ade}}}, \bibinfo {author}
  {\bibfnamefont {K.~A.}\ \bibnamefont {{Aird}}}, \bibinfo {author}
  {\bibfnamefont {J.~E.}\ \bibnamefont {{Austermann}}}, \bibinfo {author}
  {\bibfnamefont {J.~A.}\ \bibnamefont {{Beall}}}, \bibinfo {author}
  {\bibfnamefont {A.~N.}\ \bibnamefont {{Bender}}}, \bibinfo {author}
  {\bibfnamefont {B.~A.}\ \bibnamefont {{Benson}}}, \bibinfo {author}
  {\bibfnamefont {L.~E.}\ \bibnamefont {{Bleem}}}, \bibinfo {author}
  {\bibfnamefont {J.~J.}\ \bibnamefont {{Bock}}}, \bibinfo {author}
  {\bibfnamefont {J.~E.}\ \bibnamefont {{Carlstrom}}}, \bibinfo {author}
  {\bibfnamefont {C.~L.}\ \bibnamefont {{Chang}}}, \bibinfo {author}
  {\bibfnamefont {H.~C.}\ \bibnamefont {{Chiang}}}, \bibinfo {author}
  {\bibfnamefont {H.-M.}\ \bibnamefont {{Cho}}}, \bibinfo {author}
  {\bibfnamefont {A.}~\bibnamefont {{Conley}}}, \bibinfo {author}
  {\bibfnamefont {T.~M.}\ \bibnamefont {{Crawford}}}, \bibinfo {author}
  {\bibfnamefont {T.}~\bibnamefont {{de Haan}}}, \bibinfo {author}
  {\bibfnamefont {M.~A.}\ \bibnamefont {{Dobbs}}}, \bibinfo {author}
  {\bibfnamefont {W.}~\bibnamefont {{Everett}}},  \emph {et~al.},\ }\href
  {\doibase 10.1103/PhysRevLett.111.141301} {\bibfield  {journal} {\bibinfo
  {journal} {Physical Review Letters}\ }\textbf {\bibinfo {volume} {111}},\
  \bibinfo {eid} {141301} (\bibinfo {year} {2013})},\ \Eprint
  {http://arxiv.org/abs/1307.5830} {arXiv:1307.5830} \BibitemShut {NoStop}%
\bibitem [{\citenamefont {{Efstathiou}}\ and\ \citenamefont
  {{Gratton}}(2009)}]{efstathiou09}%
  \BibitemOpen
  \bibfield  {author} {\bibinfo {author} {\bibfnamefont {G.}~\bibnamefont
  {{Efstathiou}}}\ and\ \bibinfo {author} {\bibfnamefont {S.}~\bibnamefont
  {{Gratton}}},\ }\href {\doibase 10.1088/1475-7516/2009/06/011} {\bibfield
  {journal} {\bibinfo  {journal} {\jcap}\ }\textbf {\bibinfo {volume} {6}},\
  \bibinfo {eid} {011} (\bibinfo {year} {2009})},\ \Eprint
  {http://arxiv.org/abs/0903.0345} {arXiv:0903.0345 [astro-ph.CO]} \BibitemShut
  {NoStop}%
\bibitem [{\citenamefont {{Planck Collaboration C12}}(2015)}]{planck2014-a12}%
  \BibitemOpen
  \bibfield  {author} {\bibinfo {author} {\bibnamefont {{Planck Collaboration
  C12}}},\ }\href@noop {} {\bibfield  {journal} {\bibinfo  {journal} {in
  preparation}\ } (\bibinfo {year} {2015})}\BibitemShut {NoStop}%
\end{thebibliography}%

\end{document}